\documentclass[useAMS,usenatbib]{mn2e}
\pdfoutput=1
\usepackage[pdftex]{graphicx}
\usepackage{times}
\usepackage{amssymb}
\usepackage{amsmath}
\usepackage{euscript}
\usepackage{longtable,lscape}
\def\210keV{{\rm\thinspace 2--10 keV}}
\usepackage{wasysym}

\title[Simultaneous Snapshots of AGN from XMM]{Simultaneous X-ray/optical/UV snapshots of active galactic nuclei from XMM-Newton: spectral energy distributions for the reverberation mapped sample}
\author[R.V. Vasudevan and A.C. Fabian]{R.V. Vasudevan$^1$\thanks{e-mail:ranjan@ast.cam.ac.uk} and A.C. Fabian$^1$\\\footnotesize$^1$ Institute of Astronomy, Madingley Road, Cambridge CB3 0HA}

\begin{document}

\maketitle

\begin{abstract}
We employ contemporaneous optical, UV and X-ray observations from the \emph{XMM-Newton} EPIC-pn and Optical Monitor (OM) archives to present, for the first time, simultaneous spectral energy distributions (SEDs) for the majority of the Peterson et al. (2004) reverberation mapped sample of active galactic nuclei (AGN).  The raw data were reduced using the latest pipelines and are all analysed consistently.  The virial mass estimates from Peterson et al. (2004) allow us to calculate Eddington ratios $\lambda_{\rm{Edd}}$ for the sample using the bolometric accretion luminosities determined directly from the SEDs.  We calculate hard X-ray bolometric corrections $\kappa_{\rm{2-10keV}}$ for the sample and confirm a trend for increasing bolometric correction with Eddington ratio proposed in previous studies.  Our comparison with previous work on these objects suggests that the OM bandpass may be less susceptible to intrinsic reddening than the far-UV peak of the thermal disc spectrum in AGN, yielding larger bolometric corrections than previous work: $\kappa_{\rm{2-10keV}}\approx15-30$ for $\lambda_{\rm{Edd}}\lesssim0.1$, $\kappa_{\rm{2-10keV}}\approx20-70$ for $0.1\lesssim\lambda_{\rm{Edd}}\lesssim0.2$ and $\kappa_{\rm{2-10keV}}\approx70-150$ for $\lambda_{\rm{Edd}}\gtrsim0.2$, but part of this increase could be attributed to spectral complexity preventing accurate recovery of the intrinsic luminosity in some sources. Long-term optical--UV variability contributes a second-order, but significant change to the total bolometric luminosity when comparing multiple observations for individual objects.  We also consider the effect of a recently proposed correction for radiation pressure when determining black hole masses with reverberation mapping, and find that the revised mass estimates do not significantly alter the range of bolometric corrections seen but may yield a narrower distribution of Eddington ratios.

\end{abstract}


\section{Introduction}
\label{Intro}

It has long been a quest of multi-wavelength astronomy to better constrain the shape of the spectral energy distribution (SED) of active galactic nuclei (AGN), in order to better understand the nature of accretion onto supermassive black holes (SMBHs).  The established paradigm for AGN includes a thermal accretion disc emitting in the optical and ultraviolet (UV); a corona above the accretion disc emitting in X-rays by inverse compton scattering of UV disc photons; and a dusty torus which absorbs part of the UV disc emission and re-emits it in the infrared (IR).  The bulk of the accretion luminosity is visible in the optical--to--X-ray regime for most objects without heavy obscuration, and for such objects the reprocessed IR should be excluded to avoid double-counting part of the emission.

The seminal work of \cite{1994ApJS...95....1E} has provided the `template' AGN SED for many previous studies.  The mean radio-loud and radio-quiet SEDs presented in their study allow the calculation of average parameters for AGN SEDs such as bolometric corrections, which are a key requirement when estimating bolometric luminosities or accretion rates for objects. However, the spread in parameters for individual AGN is large, and average SED parameters do not reflect this variation.  \cite{2004MNRAS.351..169M} and \cite{2006astro.ph..5678H} present new `semi-empirical' SED templates which take into account the UV luminosity dependence of the optical--to--X-ray SED shape and propose a dependence of bolometric correction on luminosity.  One useful application of this luminosity dependent bolometric correction is to feed it into X-ray background synthesis models, which convert the observed energy density in X-rays from AGN into the supermassive black hole (SMBH) mass density.  Such studies are extremely illuminating in that they allow us to also constrain the accretion efficiency of AGN in the process.  However, the suggested luminosity dependence of bolometric correction does not seem to be borne out in the real AGN population.   The work of \cite{2007MNRAS.381.1235V} (`VF07' hereafter) presents observationally determined AGN SEDs for 54 AGN for which optical, far-UV and X-ray data were avialable, and present the X-ray (2--10keV) bolometric corrections $\kappa_{\rm{2-10keV}}$ ($L_{\rm{bol}}/L_{\rm{2-10keV}}$) determined from them.  In particular, they use data from the Far Ultraviolet Spectroscopic Explorer (\emph{FUSE}) which provides a window onto the peak of the accretion disc emission in the far-UV which has been historically difficult to constrain.  Their work suggests no clear luminosity trend for bolometric corrections, and highlights many uncertainties in the determination of SEDs.  Some such uncertainties include intrinsic reddening in the optical--UV part of the spectrum, variability in X-rays, lack of simultaneity in the data and uncertainties in the black hole (BH) mass estimates used to construct the SEDs.  However, they also propose a dependence of bolometric correction on Eddington ratio ($\lambda_{\rm{Edd}} = L_{\rm{bol}}/L_{\rm{Edd}}$, where $L_{\rm{Edd}}=1.38\times10^{38}(M_{\rm{BH}}/M_{\bigodot})$ for a black hole of mass $M_{\rm{BH}}$), which may now be reinforced by some studies (\citealt{2008arXiv0801.2383K}, \citealt{2008arXiv0804.0803S}).

Most AGN SEDs presented in the literature have not been constructed from contemporaneous broad-band data.  Since the degree of X-ray variability is typically far greater than the optical and UV variability, this may not be a significant uncertainty; but certainly when observations in different wavebands are separated by many months or years, the accretion state of the AGN represented in the UV disc emission is not likely to be the same as that shown by the X-ray observations.  \cite{2006MNRAS.366..953B} present a collection of SEDs for 22 Palomar Green (PG) quasars with simultaneous optical, UV and X-ray data from the \emph{XMM-Newton} satellite's EPIC-pn X-ray and Optical Monitor (OM) instruments, with a view to eliminate uncertainties in long-term variability present in non-simultaneous data.  A brief analysis of their SEDs in VF07 reveals that SED parameters determined from simultaneous SEDs may be significantly different from those from non-simultaneous ones; the comparison is made difficult since the bandpass of the OM falls just short of the canonical peak of the `big blue bump' (BBB) of accretion disc emission at $\sim1000\rm{\AA}$, which the \emph{FUSE} observations do reach.  However, using the simultaneous OM and EPIC-pn data is extremely valuable since it removes a large number of variability related uncertainties.  The OM band-pass may also be less prone to the effects of intrinsic reddening in AGN, and may therefore be useful in extrapolating a relatively `un-reddened' optical--UV SED shape for AGN.

The emergence of a possible Eddington ratio dependence of bolometric correction may provide an interesting window onto the different modes of accretion at work in different AGN.  The work of \cite{2006ApJ...646L..29S} highlights a correlation between the hard X-ray photon index $\Gamma$ and $\lambda_{\rm{Edd}}$ for AGN, reinforced in their later work \citep{2008arXiv0804.0803S}. Additionally, the Eddington ratios seen in Galactic black holes (GBH) are often known to be correlated with particular types of accretion states: the low-flux, hard states generally exhibit low values of $\lambda_{\rm{Edd}}$ whereas high-flux, softer states are typically accompanied by higher values of $\lambda_{\rm{Edd}}$ \protect\citep{2006ARA&A..44...49R}.  The potential parallels between this and the Eddington ratio dependence of AGN SED shapes emphasise the need for accurate determinations of the Eddington ratio, which require not only accurate SED data but also accurate estimates for the BH mass.  There are many different methods for determining BH mass in AGN including direct measurements using the host galaxy stellar kinematics, nuclear gas motions or the well known $M_{\rm{BH}}$--bulge luminosity correlation (\citealt{2003ApJ...589L..21M}).

One well-established technique for estimating SMBH masses is reverberation mapping (RM), which has the advantage that it does not need the high angular resolution which the above methods require.  This is a proven and powerful technique which allows the general structure and kinematics of the broad line region (BLR) in AGN to be constrained.  The foundations and assumptions behind the technique are discussed in detail in the review of \cite{Peterson:2004am}, but the essence of the approach involves monitoring the response of optical broad line fluxes from the BLR to variations in the underlying (UV) continuum flux.  The velocity dispersion $\Delta V$ estimated from the broad line-widths is related to the size $r$ of the BLR as $r \propto (\Delta V)^{-2}$.  We then require the system to be gravitationally bound, and since the gravity of the central black hole (BH) must be responsible for this, we arrive at the expression $M_{\rm{BH}}= f r (\Delta V)^{2}/G$ for the black hole mass, where $f$ is a scaling factor determined by the more detailed geometry and kinematics of the BLR.  Values for $f$ between $\sim3$ and $\sim6$ have been proposed previously, by requiring that the reverberation mass estimates obey the $M_{\rm{BH}}-\sigma$ relation (where $\sigma$ is the velocity dispersions of stars in the galaxy bulge), as done by \cite{2004ApJ...615..645O} and \cite{2008arXiv0802.2021M}.  The factor $f$ is still somewhat uncertain however, and this propagates into uncertainties of a factor of $\sim3$ in the virial mass estimates from RM.

One of the most detailed reverberation mapping studies to date is that of \cite{2004ApJ...613..682P}, who perform a thorough and consistent re-analysis of all previously available broad line reverberation mapping data on a sample of 37 AGN, finding BH mass estimates for 35 of them (table 8 of their work).  The mass estimates for two AGN in their sample, NGC 4593 and IC 4329A have extremely large error bars, but in the case of NGC 4593 a more recent, better constrained RM mass estimate is available from \cite{2006ApJ...653..152D}, calculated using a newer determination of the time lag between the BLR and continuum. SEDs for 21 of the AGN covered by \cite{2004ApJ...613..682P} were presented in VF07 but using older mass estimates from diverse sources in the literature and non-contemporaneous data.  In this work we refine the approach employed in VF07 and present newly determined SEDs for a total of 29 objects from the \cite{2004ApJ...613..682P} sample.

Since the RM technique requires that the optical--UV spectra are monitored for a sufficient time period to be able to detect changes in continuum emission, this necessitates that the sample is sufficiently bright in the optical and UV.  We therefore miss out on optically faint classes of AGN such as obscured/type II AGN, but a detailed investigation of the accretion emission in obscured objects is beyond the scope of this study.  We instead make the most of the \cite{2004ApJ...613..682P} sample (augmented with the RM mass for NGC 4593 from \citealt{2006ApJ...653..152D}) as providing a set of mass estimates uniformly determined by one method, and combine this with simultaneous optical, UV and X-ray data from the \emph{XMM-Newton} archives as done by \cite{2006MNRAS.366..953B}.  This allows us to construct simultaneous SEDs and calculate Eddington fractions and bolometric corrections with significantly reduced systematic uncertainties due to long-term variability and discrepancies between mass estimation techniques.  This work represents the first such determination of simultaneous SEDs for AGN with consistent mass measurements.

\section{Data Sources and Reduction}

The simultaneous multi-wavelength data on the AGN in the \cite{2004ApJ...613..682P} sample were obtained from the XMM pn and OM instruments (with the MOS data only being used in the initial stages for source detection purposes).  Usable data were not available for five AGN: PG 0026+129, Mrk 817, P 1617+175, PG 1700+518 (for which no \emph{XMM} data were available) and PG 0804+761 (for which no OM data was available in the one available \emph{XMM} observation).  The original data files (`ODFs') for the observations were downloaded from the XMM-Newton Science Archive (XSA)\footnote{http://xmm.esac.esa.int/external/xmm\_data\_acc/xsa/index.shtml}. We generally follow the approach of \cite{2006MNRAS.366..953B} in reducing the data. As part of our initial checks, we downloaded a few ODFs used in their work and ensured that the main results (such as photon indices and luminosities) were reproduced by our approach, within errors and allowing for changes in the XMM Science Analysis System (SAS) pipelines. A list of the observation ID numbers for the observations used is given in Table \ref{table:datadetails} below, along with details of the dates, observation times and window modes used for each observation.  

The X-ray data were reduced using the SAS v7.1.0 \textsc{epchain} and \textsc{emchain} for the pn and MOS instruments respectively. Automatic source-detection was performed using the \textsc{edetect\_chain} routines where possible; if this routine failed for a particular source, the source and background regions were identified manually. Circular source regions (default radius 36 arcsec) were used to produce source event files and source spectra.  The pn only was used for all subsequent parts of the analysis, and the pn event file was filtered as specified according to the standard guidelines (allowing only photons with pattern 0--4 and quality flag 0).  Source-free regions near the source - preferentially on the same CCD chip - were used to extract background event files and spectra. The datasets were inspected for pile-up using the \textsc{epatplot} routine and the 10--12 keV background light curves were inspected for flaring.  Annular source regions were used in sources where pile-up was detected, and the \textsc{tabgtigen} tool was used to set a count-rate threshold from which to calculate the usable time intervals in each observation.  The results of these checks are also recorded in Table \ref{table:datadetails}.  Response matrices and auxiliary files were generated using the \textsc{rmfgen} and \textsc{arfgen} tools, and the final pn spectra were grouped with a minimum of 20 counts per bin using the \textsc{grppha} tool.

The Optical and UV data from the OM were reduced using the \textsc{omichain} pipeline.  Point source and extended source identification is automatically performed as part of this pipeline, and the point source corresponding to the nucleus was selected using the images generated for each waveband, ensuring that the identified optical or UV counterpart was coincident with the source region used for the X-ray spectrum.  Since the reverberation-mapped sample will be relatively bright by necessity, we do not make any specific attempt to correct for host-galaxy contamination to the nuclear magnitudes obtained, and in most cases, the nucleus was clearly identifiable in the OM images.  The optical--UV photometry results were extracted from the \textsc{swsrli} results files, and the magnitudes corrected for Galactic extinction using the extinction law of \cite{1989ApJ...345..245C}.  Values for the Galactic extinction $E(B-V)$ were obtained from the NASA/IPAC Infrared Science Archive\footnote{http://irsa.ipac.caltech.edu/applications/DUST/}.  The data points from the OM were converted to \textsc{xspec} spectrum files using the \textsc{flx2xsp} utility, part of the \textsc{ftools} package.

\begin{table*}
\begin{tabular}{|p{2cm}|p{1.3cm}|p{1.8cm}|p{1.5cm}|p{2cm}|p{1.4cm}|p{2.3cm}|p{1.5cm}|l}
\hline
Object&Redshift&XMM\newline Observation ID&Observation\newline date&Window mode$^{1}$&Observation time (ks)&Usable percentage of observation time&Source Variability ($\mathrm{{\Delta}I_{rms}/{\langle}I{\rangle}}$) $^{2}$&Pile-up$^{3}$\\\hline
3C 120&0.03301&0152840101&2003-08-26&SW&133.8&95.6&0.0857&---\\
3C 390.3&0.05610&0203720201&2004-10-08&SW&70.4&84.7&0.0833&---\\
Ark 120&0.03271&0147190101&2003-08-24&SW&112.1&94.8&0.0700&---\\
Fairall 9&0.04702&0101040201&2000-07-05&FF&33.0&89.6&0.3835&Y\\
IC 4329A&0.01605&0147440101&2003-08-06&SW&136.0&86.0&0.0727&---\\
Mrk 110&0.03529&0201130501&2004-11-15&SW&47.4&94.4&0.0864&---\\
Mrk 279 (1)&0.03045&0302480601&2005-11-19&SW&38.2&88.0&0.1925&---\\
Mrk 279 (2)&0.03045&0302480401&2005-11-15&SW&59.8&83.7&0.0875&---\\
Mrk 279 (3)&0.03045&0302480501&2005-11-17&SW&59.8&87.5&0.1548&---\\
Mrk 335 (1)&0.02578&0510010701&2007-07-10&LW&22.6&91.7&0.2333&---\\
Mrk 335 (2)&0.02578&0101040101&2000-12-25&FF&36.9&88.2&0.3220&Y\\
Mrk 509&0.03440&0306090401&2006-04-25&SW&70.0&92.3&0.0794&---\\
Mrk 590&0.02638&0201020201&2004-07-04&SW&112.7&86.1&0.5200&---\\
Mrk 79&0.02219&0103860801&2000-10-09&SW&2.5&88.6&0.2211&---\\
NGC 3227 (1)&0.00386&0400270101&2006-12-03&LW&107.9&99.2&0.1235&---\\
NGC 3227 (2)&0.00386&0101040301&2000-11-28&FF&40.1&99.1&0.1982&---\\
NGC 3516&0.00884&0107460701&2001-11-09&SW&130.0&94.9&0.1715&---\\
NGC 3783 (1)&0.00973&0112210501&2001-12-19&SW&137.8&96.4&0.1306&---\\
NGC 3783 (2)&0.00973&0112210201&2001-12-17&SW&137.8&99.8&0.1425&---\\
NGC 4051 (1)&0.00234&0109141401&2001-05-16&SW&122.0&90.6&0.3968&---\\
NGC 4051 (2)&0.00234&0157560101&2002-11-22&LW&51.9&98.2&0.7795&Y\\
NGC 4151 (1)&0.00332&0143500101&2003-05-25&SW&19.0&86.1&0.0839&---\\
NGC 4151 (2)&0.00332&0143500201&2003-05-26&SW&18.9&99.2&0.0874&---\\
NGC 4151 (3)&0.00332&0143500301&2003-05-27&SW&18.9&98.8&0.0929&---\\
NGC 4593*&0.00900&0109970101&2000-07-02&SW&28.1&78.4&0.1036&---\\
NGC 5548&0.01717&0109960101&2000-12-24&SW&26.1&92.1&0.0965&---\\
NGC 7469&0.01632&0112170101&2000-12-26&SW&19.0&94.3&0.1309&---\\
PG 0052+251&0.15500&0301450401&2005-06-26&SW&20.5&84.3&0.1795&---\\
PG 0844+349&0.06400&0103660201&2000-11-04&FF&26.4&89.6&0.4297&Y\\
PG 0953+414&0.23410&0111290201&2001-11-22&LW&15.6&98.7&0.1568&---\\
PG 1211+143 (1)&0.08090&0208020101&2004-06-21&LW&60.0&83.8&0.3253&Y\\
PG 1211+143 (2)&0.08090&0112610101&2001-06-15&LW&55.7&84.2&0.4841&Y\\
PG 1226+023&0.15834&0414190101&2007-01-12&SW&78.6&94.5&0.0435&---\\
PG 1229+204&0.06301&0301450201&2005-07-09&SW&25.5&100.0&0.4125&---\\
PG 1307+085&0.15500&0110950401&2002-06-13&LW&14.0&94.1&0.2914&---\\
PG 1411+442&0.08960&0103660101&2002-07-10&EF&41.8&85.1&1.3188&---\\
PG 1426+015&0.08647&0102040501&2000-07-28&FF&17.6&93.5&0.3704&Y\\
PG 1613+658&0.12900&0102040601&2001-04-13&FF&12.8&67.5&0.4279&---\\
PG 2130+099&0.06298&0150470701&2003-05-16&SW&37.9&91.3&0.2766&---\\
\hline
\end{tabular}
\caption{\label{table:datadetails}Details of XMM observations used in this paper. 1 -- SW: Small Window mode, FF: Full Frame mode, LW: Large Window mode, EF: Extended Full Frame mode. 2 -- Variability recorded in the rest frame 2--5 keV band. 3 -- Sources marked `Y' were corrected for pile-up as detailed in the text. * For NGC 4593, the better constrained $M_{\rm{BH}}$ estimate from \protect\cite{2006ApJ...653..152D} was used. \protect\cite{2004ApJ...613..682P} $M_{\rm{BH}}$ estimates were used for the rest of the sample.}
\end{table*}

\section{Construction of SEDs}
\label{modelfitting}

The \textsc{xspec12} package was used to fit a basic accretion disk and absorbed power-law model (\textsc{diskpn+wabs(zwabs(bknpower))}) to the reduced pn and OM data.  The \textsc{diskpn} model takes the parameters $T_{\rm{max}}$ (maximum temperature in the accretion disc), $R_{\rm{in}}$ (inner radius of the accretion disc) and $K$ (overall normalisation).  The inner radius $R_{in}$ was frozen at 6.0 gravitational radii, and the temperature $T_{\rm{max}}$ was left as a free parameter.  The normalisation for the \textsc{diskpn} model is defined as $K = M_{\rm{BH}}^2 cos(i)/D_{\rm{L}}^2\beta^4$, where $M_{\rm{BH}}$ is the black hole mass, $i$ the inclination angle, $D_{\rm{L}}$ the luminosity distance and $\beta$ the colour-to-effective temperature ratio. SMBH mass estimates were taken from \cite{2004ApJ...613..682P} (or \citealt{2006ApJ...653..152D}, in the case of NGC 4593), and luminosity distances from Ned Wright's Cosmology Calculator (assuming a cosmology of $H_0 = 71$ $\mathrm{kms^{-1}Mpc^{-1}}$, $\Omega_{\rm{M}}=0.27$).  We performed brief tests to ascertain the importance of inclination angle. Adopting typical values between $0^\circ$ and $35^\circ$ can produce changes in $K$ of up to 18 per cent, but the total bolometric luminosity is a function of not only $K$ but also $T_{\rm{max}}$ and the spectral fit in the X-ray regime.  Despite the 18 per cent change in $K$ due to inclination angle, this propagates to give changes of only $\sim$6 per cent in the bolometric luminosities calculated from the SEDs, which was generally smaller than the statistical error in the luminosities themselves.  Therefore, a value of $i=0^\circ$ was adopted in all cases, and $\beta$ was assumed to be Unity.  The normalisation thus determined was kept frozen.

\begin{figure}
    \includegraphics[width=8.5cm]{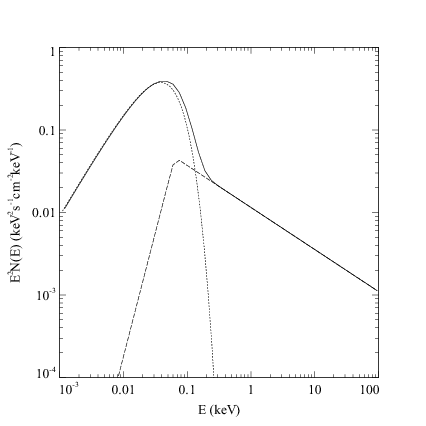}
    \caption{An illustration of the model combination used (wihout intrinsic or Galactic absorption components).  The dotted line represents the \textsc{diskpn} model, the dashed line represents the \textsc{bknpower} model and the solid line shows the resultant total model.  The \textsc{bknpower} model was employed to ensure that the X-ray power law was not extrapolated to UV and optical energies, below the Big Blue Bump energy range.  The power law index for the low-energy branch of the broken power law was set to be -1.}
\label{examplemodel}
\end{figure}

The \textsc{bknpower} model was employed to fit the X-ray data, with the low energy branch of the model set to have negligible values within the energy range of the BBB.  This ensured that the power law was not extrapolated to give spurious fluxes at energies below the thermal disc spectrum; this is a particular danger for sources with soft ($\Gamma<2$) hard X-ray spectra (see Fig. \ref{examplemodel}).  The break energy between the two regimes was set to be 4 times the \textsc{diskpn} $T_{\rm{max}}$ parameter.  The \textsc{wabs} model was used to incorporate Galactic absorption in the X-ray regime, with the Galactic neutral Hydrogen column density, $N_{H}^{(Gal)}$ frozen at the value provided by the \textsc{nh} ftool.  The \textsc{zwabs} component allowed for some element of absorption intrinsic to the source and the intrinsic column density was left as a free parameter. Bad channels in the X-ray spectrum were ignored, and only the 1--8 keV (observed frame) data were used for fitting the X-ray continuum, to avoid the soft excess and any features above 10 keV such as the reflection hump.  Because of the simple approach adopted for modelling the continuum here, we do not use the values of $N_{H}$ from the resultant fits for any further analysis, and suggest that for many objects with complex absorption, the far more detailed fits available in the literature should be referred to for a more accurate determination of their intrinsic column densities.

\begin{figure*}
    \includegraphics[width=5cm]{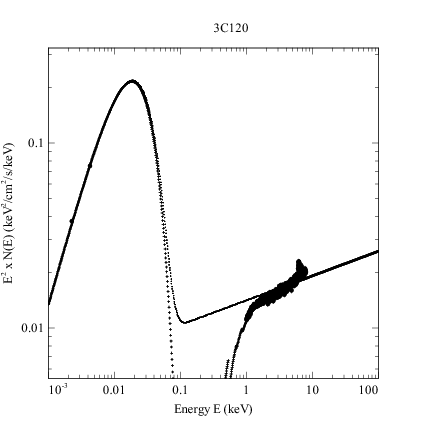}
    \includegraphics[width=5cm]{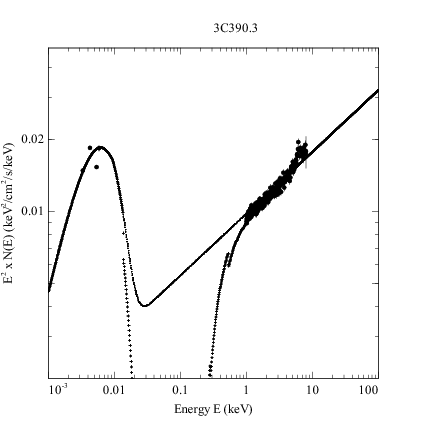}
    \includegraphics[width=5cm]{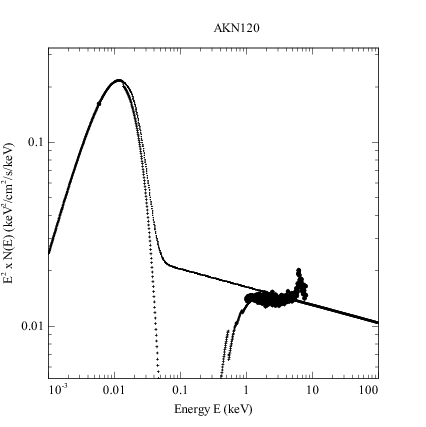}
    \includegraphics[width=5cm]{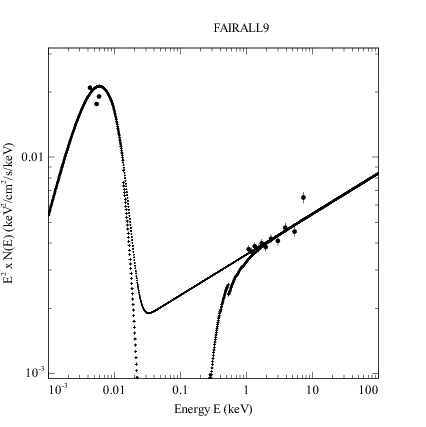}
    \includegraphics[width=5cm]{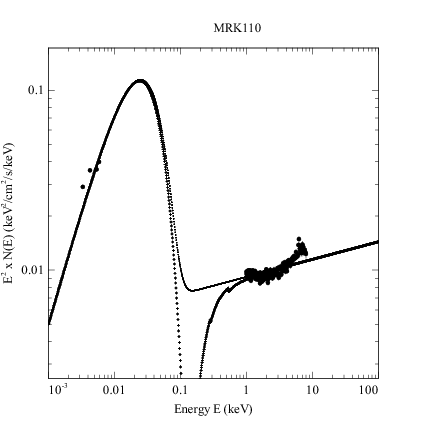}
    \includegraphics[width=5cm]{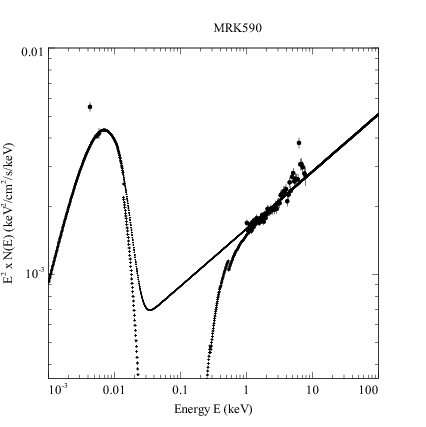}
    \includegraphics[width=5cm]{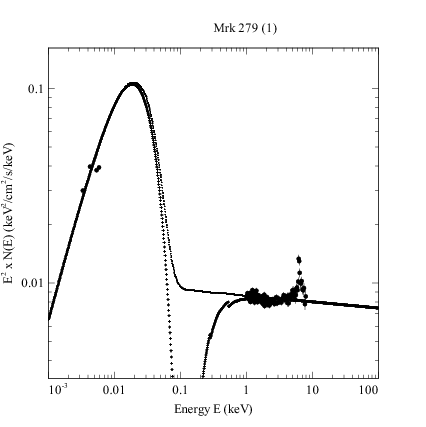}
    \includegraphics[width=5cm]{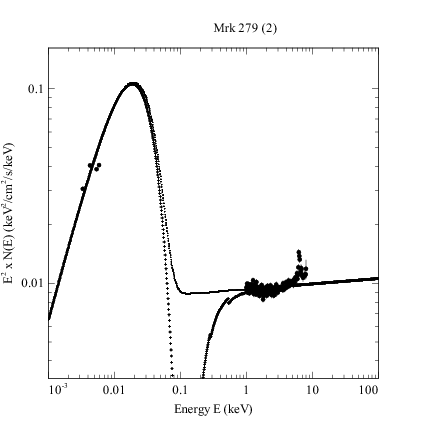}
    \includegraphics[width=5cm]{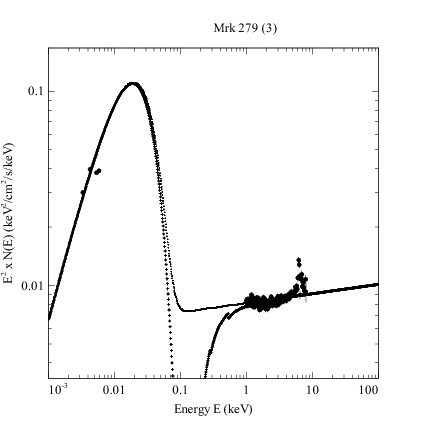}
    \includegraphics[width=5cm]{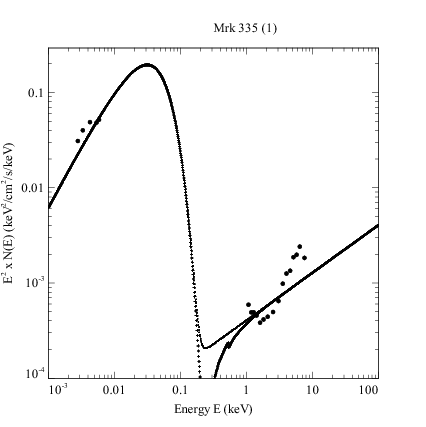}
    \includegraphics[width=5cm]{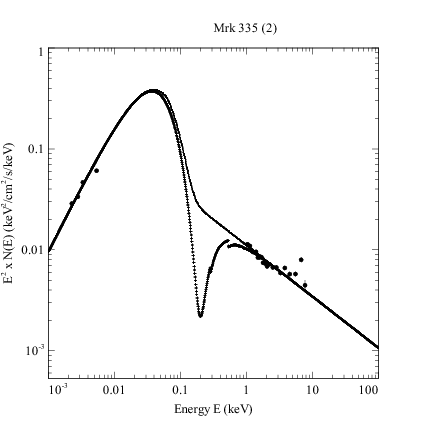}
    \includegraphics[width=5cm]{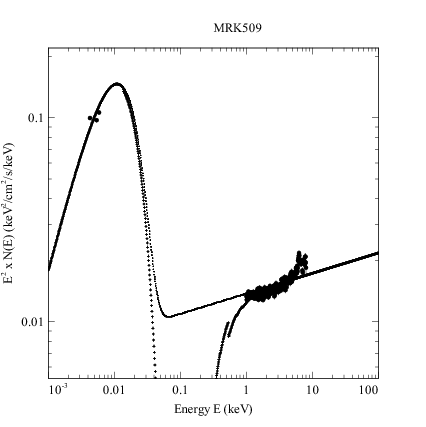}
  \caption{Spectral Energy Distributions for the AGN in the \protect\cite{2004ApJ...613..682P} sample. Filled circles represent the XMM pn and OM data, crosses represent the model fit with absorption and the dotted line represents the model with absorption removed.}
\label{seds}
\end{figure*}

  We inspected the resultant fits, and in cases where this simple model yielded a best fit photon index $\Gamma$ smaller than 1.5, we fixed $\Gamma$ at 1.5 in a revised version of the fit on the grounds that such low photon indices would be unphysical in canonical inverse-Compton scattering models for the coronal emission.  This was done for the objects NGC 3227, NGC 3516, NGC 3783, NGC 4151, PG 1411+442, Mrk 335 (1) and PG 1307+085.  The choice of $\Gamma = 1.5$ is broadly consistent with the values found in more detailed studies on these particular observations (\protect\citealt{2003A&A...397..883G} for NGC 3227, \citealt{2005ApJ...618..155T} for NGC 3516 and \citealt{2004ApJ...602..648R} for NGC 3783).   We present the resultant SEDs for 29 objects with plausible fits to the data in Fig. \ref{seds} (IC 4329A did not yield a satisfactory fit in the optical--UV).  We calculate unabsorbed luminosities in the 2--10keV, 0.001--100keV (bolometric) bands using a modified version of the `fluxerror.tcl' code provided on the \textsc{heasarc} website\footnote{http://heasarc.nasa.gov/xanadu/xspec/fluxerror.html} for this purpose.  Monochromatic unabsorbed luminosities were determined at $2500\rm{\AA}$ and $2$ keV for determining the optical--to--X-ray spectral slope, $\alpha_{\rm{OX}}$.  The calculated values for $L_{\rm{2-10keV}}$, $L_{\rm{bol}}$, $\lambda_{\rm{Edd}}$, $\kappa_{\rm{2-10keV}}$ and $\alpha_{\rm{OX}}$ for each of these objects is presented in Table \ref{table:results}.

\begin{figure*}
    \includegraphics[width=5cm]{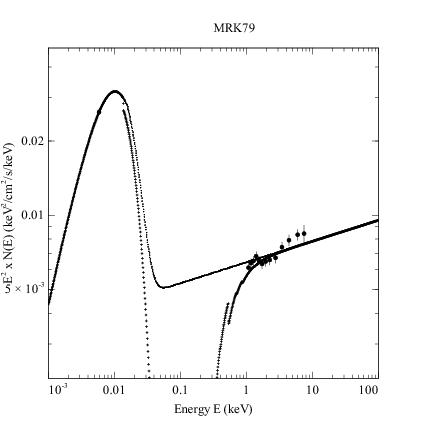}
    \includegraphics[width=5cm]{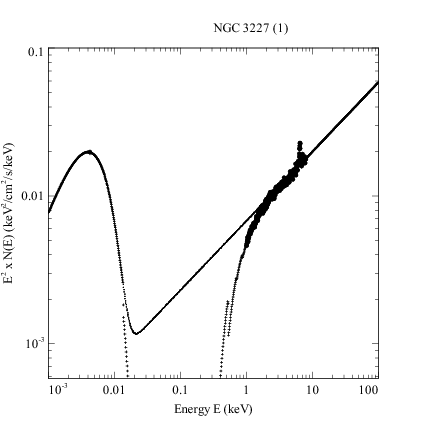}
    \includegraphics[width=5cm]{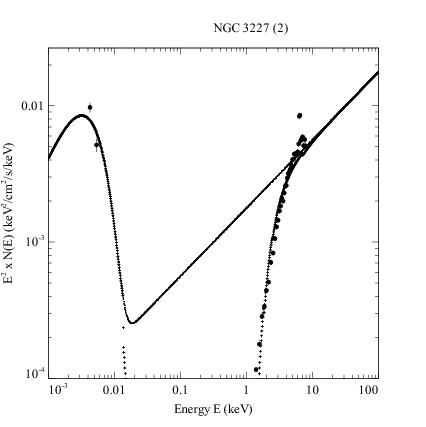}
    \includegraphics[width=5cm]{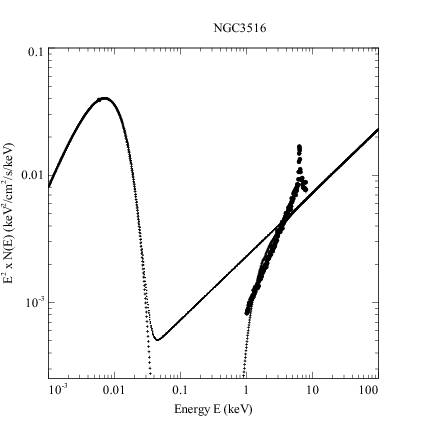}
    \includegraphics[width=5cm]{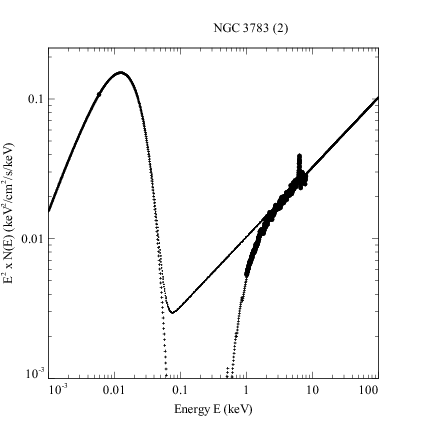}
    \includegraphics[width=5cm]{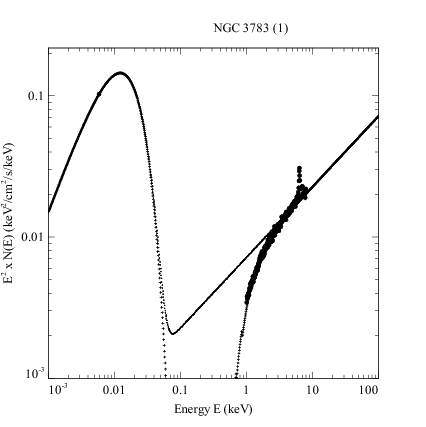}
    \includegraphics[width=5cm]{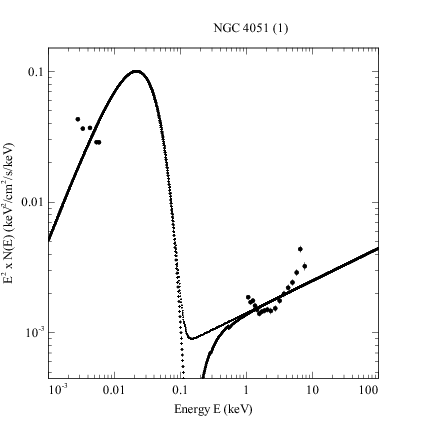}
    \includegraphics[width=5cm]{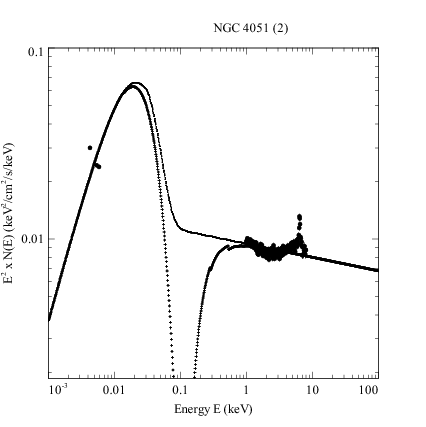}
    \includegraphics[width=5cm]{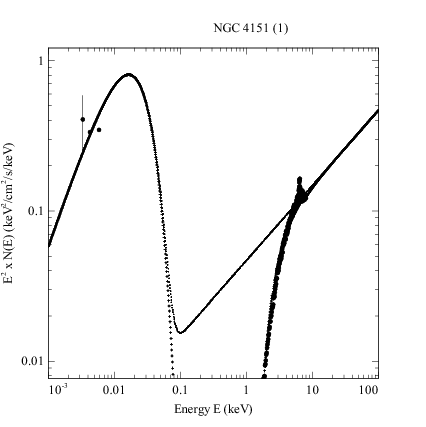}
    \includegraphics[width=5cm]{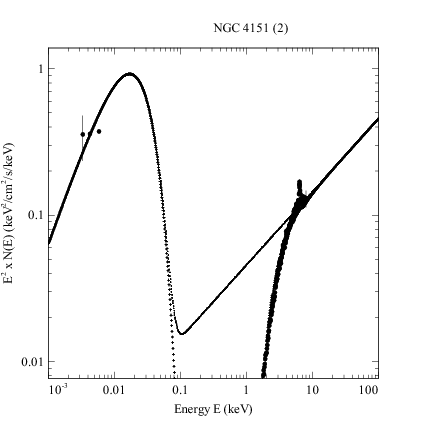}
    \includegraphics[width=5cm]{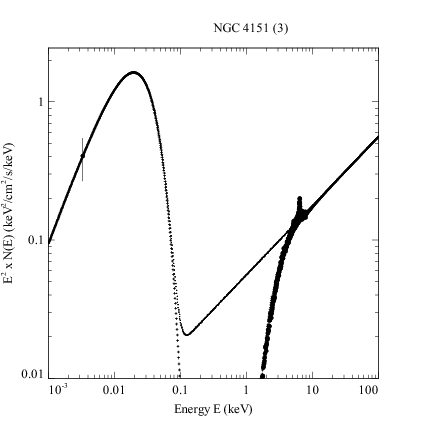}
    \includegraphics[width=5cm]{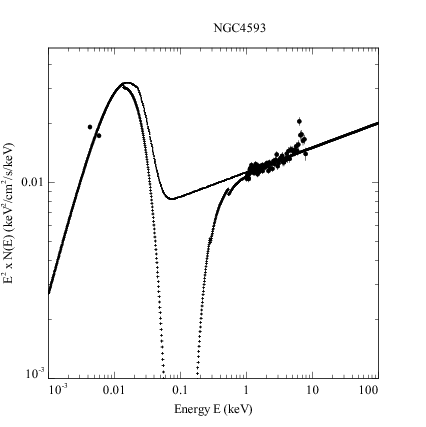}
    \begin{center}
      Figure~\ref{seds} (continued)
    \end{center}
\end{figure*}
\begin{figure*}
    \includegraphics[width=5cm]{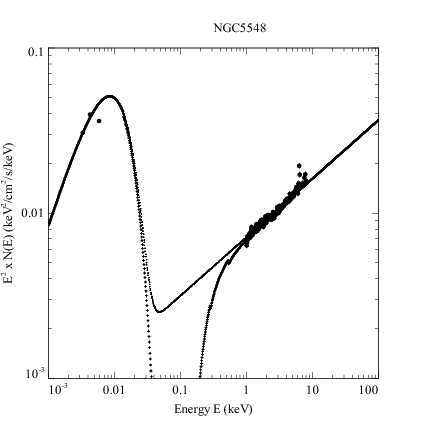}
    \includegraphics[width=5cm]{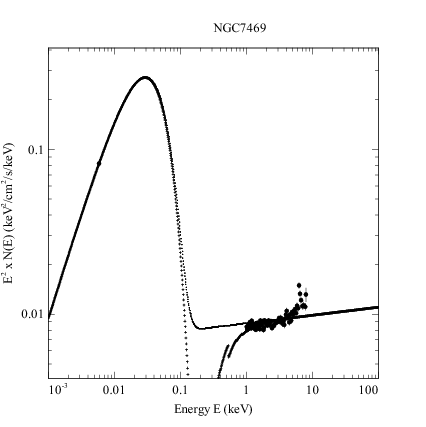}
    \includegraphics[width=5cm]{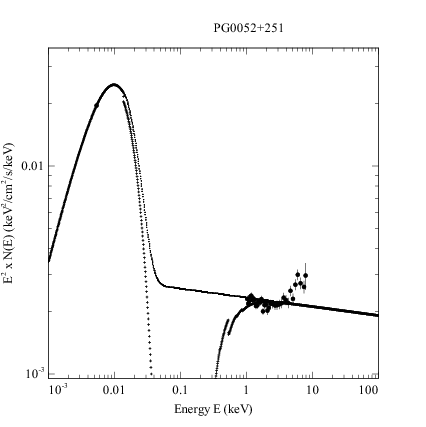}
    \includegraphics[width=5cm]{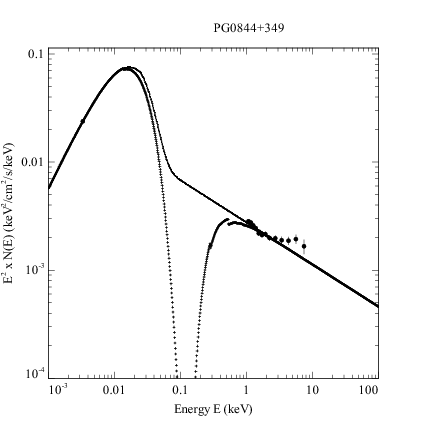}
    \includegraphics[width=5cm]{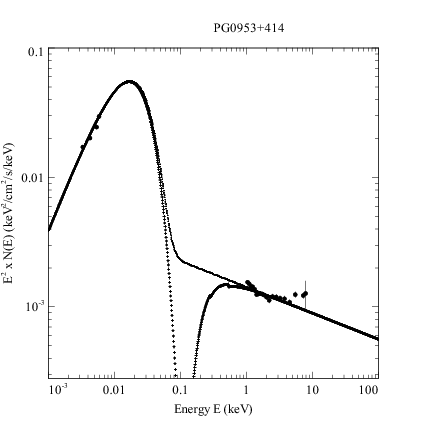}
    \includegraphics[width=5cm]{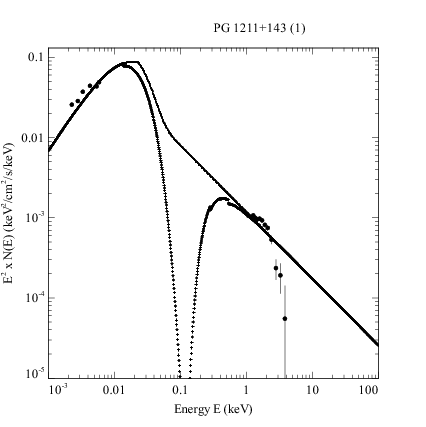}
    \includegraphics[width=5cm]{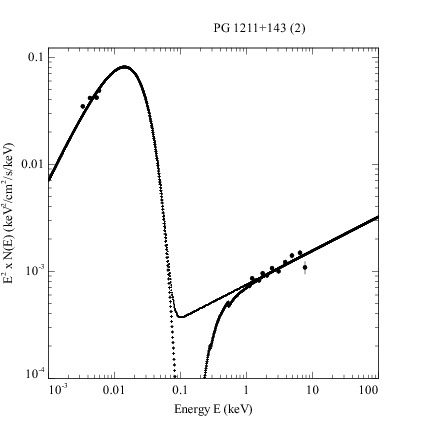}
    \includegraphics[width=5cm]{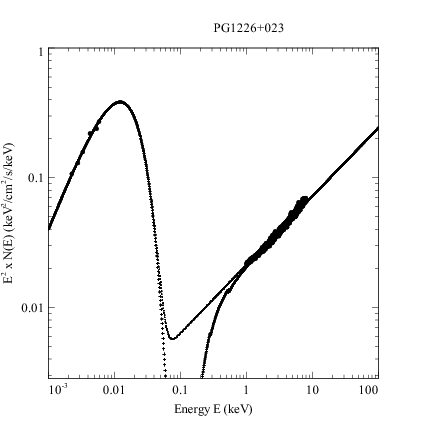}
    \includegraphics[width=5cm]{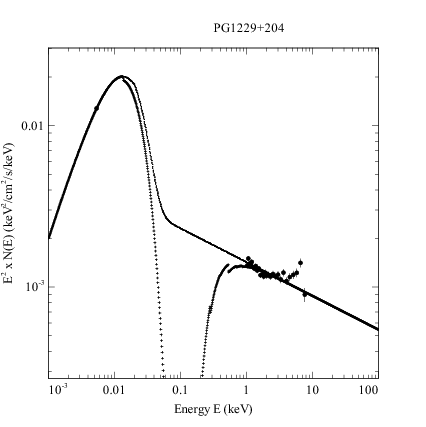}
    \includegraphics[width=5cm]{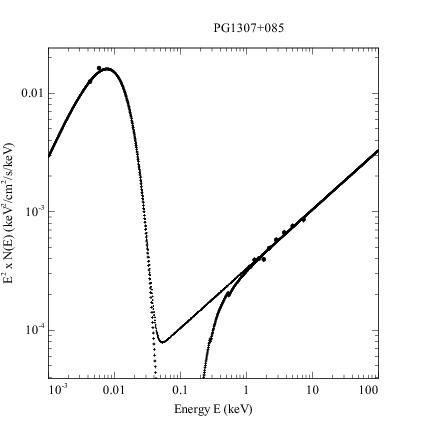}
    \includegraphics[width=5cm]{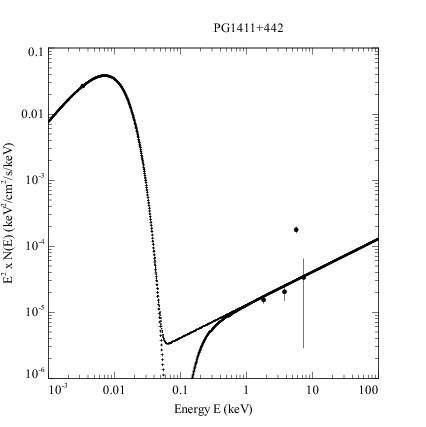}
    \includegraphics[width=5cm]{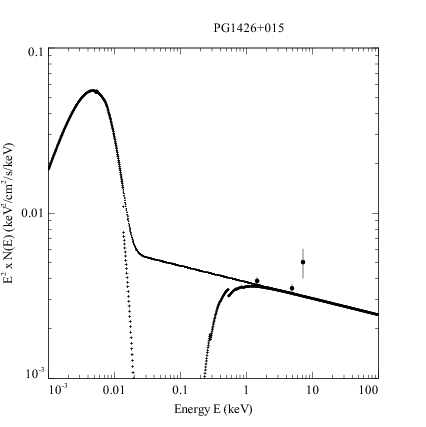}
    \begin{center}
      Figure~\ref{seds} (continued)
    \end{center}
\end{figure*}

\begin{figure*}
    \includegraphics[width=5cm]{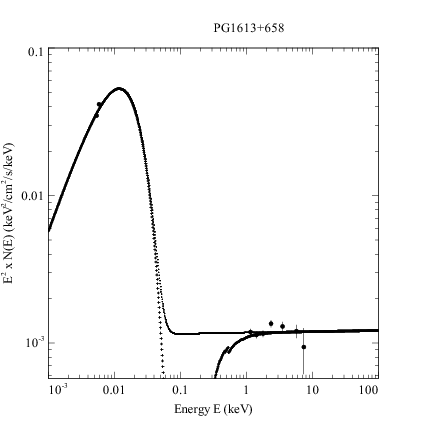}
    \includegraphics[width=5cm]{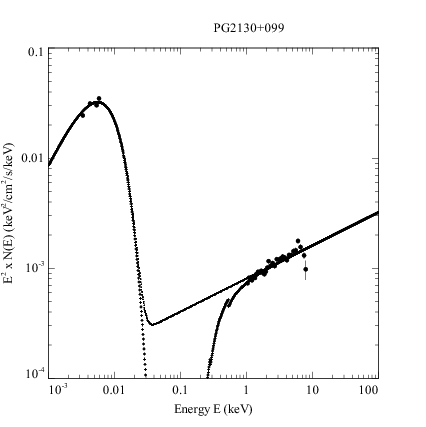}
    \begin{center}
      Figure~\ref{seds} (continued)
    \end{center}
\end{figure*}

\begin{table*}
\begin{tabular}{|p{2.0cm}|p{1.5cm}|p{1.6cm}|p{2.6cm}|p{2.3cm}|p{1.5cm}|p{1.5cm}|p{1.8cm}}
\hline
AGN&log($M_{\rm{BH}}/M_{\rm{\bigodot}}$)&Photon index \newline $\Gamma$&log({$L_{\rm{2-10keV}}/\rm{erg s^{-1}}$})&log({$L_{\rm{bol}}/\rm{erg s^{-1}}$})&Eddington ratio\newline$\lambda_{\rm{Edd}}$&Bolometric\newline correction\newline$\kappa_{\rm{2-10keV}}$&$\alpha_{\rm{OX}}$\\
\hline
3C120&$7.74$&$1.87$&$44.0$&$45.3$&$0.305$&$20.6^{+0.1}_{-0.1}$&$-1.29$\\
3C390.3&$8.46$&$1.74$&$44.4$&$45.2$&$0.0466$&$6.33^{+0.04}_{-0.03}$&$-1.07$\\
AKN120&$8.18$&$2.10$&$43.9$&$45.3$&$0.111$&$25.0^{+0.2}_{-0.3}$&$-1.37$\\
FAIRALL9&$8.41$&$1.81$&$43.8$&$44.8$&$0.0186$&$10.5^{+0.8}_{-0.8}$&$-1.28$\\
MRK110&$7.40$&$1.90$&$43.9$&$45.1$&$0.433$&$18.4^{+0.1}_{-0.1}$&$-1.21$\\
MRK279 (1)&$7.54$&$2.03$&$43.6$&$45.0$&$0.210$&$21.7^{+0.3}_{-0.3}$&$-1.27$\\
MRK279 (2)&$7.54$&$1.97$&$43.7$&$45.0$&$0.216$&$18.9^{+0.1}_{-0.1}$&$-1.25$\\
MRK279 (3)&$7.54$&$1.95$&$43.7$&$45.0$&$0.214$&$20.9^{+0.1}_{-0.1}$&$-1.27$\\
MRK335 (1)&$7.15$&$1.50*$&$42.5$&$45.0$&$0.515$&$292^{+9}_{-8}$&$-1.72$\\
MRK335 (2)&$7.15$&$2.51$&$43.3$&$45.3$&$1.13$&$102^{+4}_{-4}$&$-1.36$\\
MRK509&$8.16$&$1.90$&$44.0$&$45.2$&$0.0951$&$16.20^{+0.10}_{-0.09}$&$-1.32$\\
MRK590&$7.68$&$1.75$&$43.0$&$43.8$&$0.0104$&$7.0^{+0.2}_{-0.2}$&$-1.12$\\
MRK79&$7.72$&$1.91$&$43.3$&$44.3$&$0.0309$&$10.5^{+1.0}_{-0.9}$&$-1.20$\\
NGC3227 (1)&$7.63$&$1.53$&$42.1$&$42.9$&$0.00151$&$7.02^{+0.05}_{-0.05}$&$-1.12$\\
NGC3227 (2)&$7.63$&$1.50*$&$41.5$&$42.4$&$0.000447$&$8.0^{+0.1}_{-0.1}$&$-1.17$\\
NGC3516&$7.63$&$1.50*$&$42.3$&$43.5$&$0.00612$&$15.32^{+0.10}_{-0.10}$&$-1.40$\\
NGC3783 (1)&$7.47$&$1.50*$&$43.1$&$44.2$&$0.0433$&$14.28^{+0.10}_{-0.10}$&$-1.31$\\
NGC3783 (2)&$7.47$&$1.50*$&$42.9$&$44.1$&$0.0363$&$17.3^{+0.2}_{-0.2}$&$-1.36$\\
NGC4051 (1)&$6.28$&$1.75$&$40.8$&$42.6$&$0.0164$&$67^{+4}_{-3}$&$-1.51$\\
NGC4051 (2)&$6.28$&$2.07$&$41.4$&$42.6$&$0.0151$&$15.1^{+0.2}_{-0.1}$&$-1.17$\\
NGC4151 (1)&$7.12$&$1.50*$&$42.8$&$44.0$&$0.0558$&$15.64^{+0.08}_{-0.08}$&$-1.29$\\
NGC4151 (2)&$7.12$&$1.50*$&$42.8$&$44.0$&$0.0616$&$17.38^{+0.08}_{-0.08}$&$-1.31$\\
NGC4151 (3)&$7.12$&$1.50*$&$42.9$&$44.2$&$0.0904$&$22^{+9}_{-9}$&$-1.34$\\
NGC4593&$6.99\dagger$&$1.87$&$42.8$&$43.7$&$0.0369$&$7.7^{+0.1}_{-0.1}$&$-1.05$\\
NGC5548&$7.83$&$1.65$&$43.3$&$44.3$&$0.0236$&$10.1^{+0.1}_{-0.1}$&$-1.25$\\
NGC7469&$7.09$&$1.95$&$43.2$&$44.8$&$0.369$&$42^{+1}_{-1}$&$-1.33$\\
PG0052+251&$8.57$&$2.04$&$44.6$&$45.8$&$0.148$&$19.5^{+0.7}_{-0.6}$&$-1.33$\\
PG0844+349&$7.97$&$2.39$&$43.6$&$45.4$&$0.233$&$72^{+8}_{-8}$&$-1.46$\\
PG0953+414&$8.44$&$2.20$&$44.7$&$46.5$&$0.892$&$71^{+2}_{-2}$&$-1.46$\\
PG1211+143 (1)&$8.16$&$2.83$&$43.2$&$45.7$&$0.260$&$340^{+80}_{-60}$&$-1.68$\\
PG1211+143 (2)&$8.16$&$1.68$&$43.7$&$45.6$&$0.224$&$92^{+5}_{-5}$&$-1.64$\\
PG1226+023&$8.95$&$1.47$&$45.9$&$47.1$&$1.14$&$16.53^{+0.05}_{-0.05}$&$-1.33$\\
PG1229+204&$7.86$&$2.21$&$43.4$&$44.9$&$0.0823$&$31^{+2}_{-2}$&$-1.37$\\
PG1307+085&$8.64$&$1.50*$&$44.0$&$45.6$&$0.0659$&$35^{+1}_{-1}$&$-1.56$\\
PG 1411+442&$8.65$&$1.50*$&$42.1$&$45.4$&$0.0414$&$1700^{+800}_{-400}$&$-2.26$\\
PG1426+015&$9.11$&$2.10$&$44.1$&$45.6$&$0.0243$&$30^{+40}_{-10}$&$-1.48$\\
PG1613+658&$8.45$&$1.99$&$44.1$&$45.9$&$0.221$&$67^{+20}_{-10}$&$-1.55$\\
PG2130+099&$8.66$&$1.70$&$43.5$&$45.0$&$0.0179$&$35^{+1}_{-1}$&$-1.58$\\
\hline
\end{tabular}
\caption{\label{table:results}Luminosities, accretion rates, hard X-ray bolometric corrections and $\alpha_{\rm{OX}}$ values for 29 AGN from the \protect\cite{2004ApJ...613..682P} sample. $\dagger$ - The revised mass estimate from \protect\cite{2006ApJ...653..152D} was used for NGC 4593. Objects marked with an asterisk (*) had the photon index fixed at $\Gamma=1.5$ (see \S \ref{modelfitting} for details).  Errors are omitted for all quantities (except bolometric corrections) as they are generally small and systematic uncertainties are expected to dominate.}
\end{table*}

\section{Discussion}

\subsection{Checking the $\mathrm{\alpha_{OX}-L_{\nu}(2500\AA)}$ relation}

\begin{figure}
    \includegraphics[width=8.5cm]{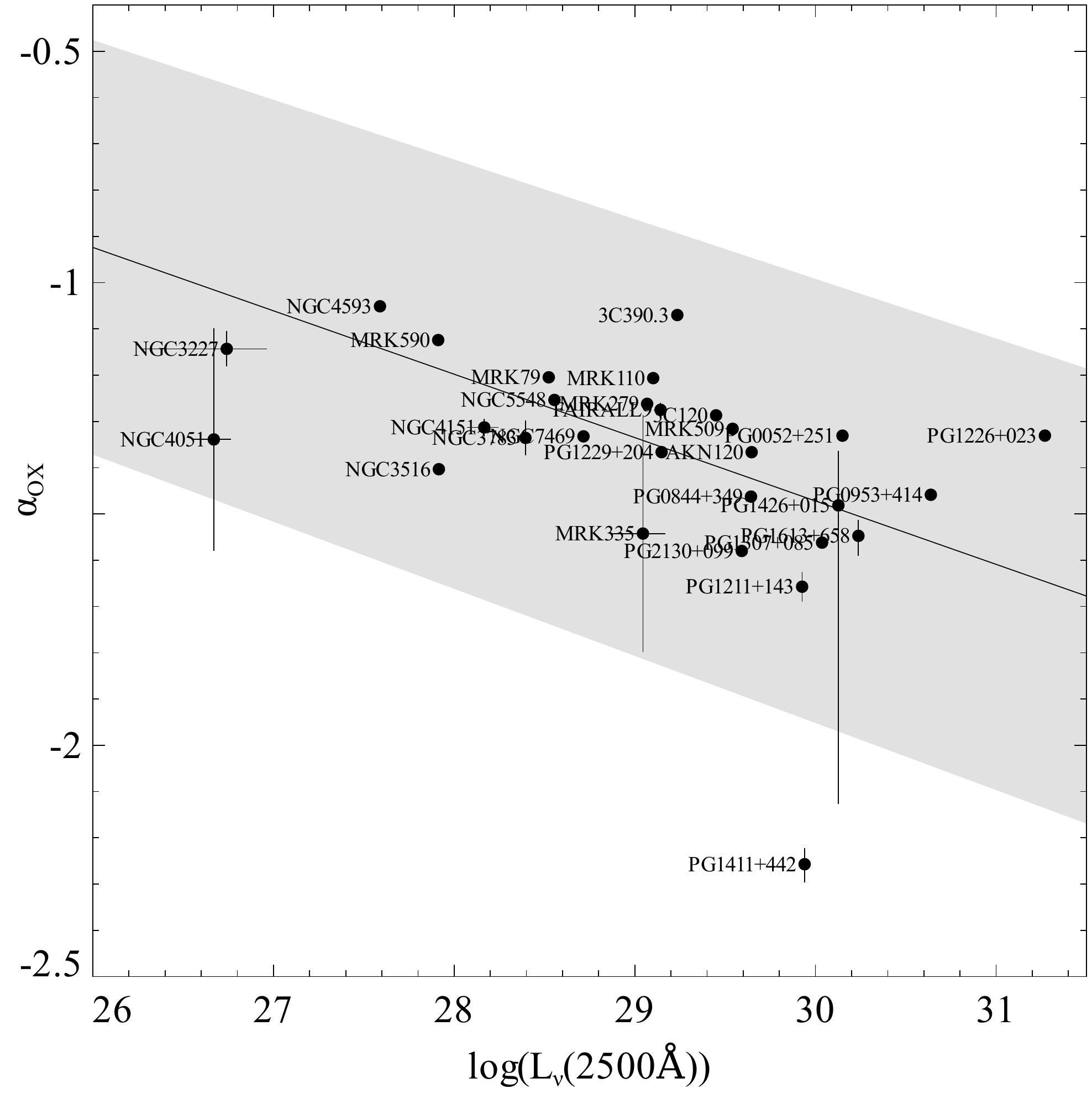}
    \caption{X-ray loudness $\mathrm{\alpha_{OX}}$ against $\mathrm{L_{\nu}(2500\AA)}$. Filled circles represent the $\mathrm{\alpha_{OX}}$ values determined from our sample; the black line is the best fit from \protect\cite{2006AJ....131.2826S} and the grey shaded area shows the spread in their best fit.}
\label{alphaox}
\end{figure}

As in VF07, we check the SEDs generated against the $\mathrm{\alpha_{OX}-L_{\nu}(2500\AA)}$ relation from the literature.  It is known from the literature that the near-UV--to--X-ray spectral slope varies with the near-UV luminosity, and we expect our SEDs to be consistent with previous determinations of this relation.  For definitions of $\mathrm{\alpha_{OX}}$ and associated quantities, please refer to \cite{2005AJ....130..387S}.  The results are shown in Fig. \ref{alphaox}, along with the recent best-fit determination of this relation from \cite{2006AJ....131.2826S}.  The calculated values of $\alpha_{\rm{OX}}$ lie within the scatter expected from the \cite{2006AJ....131.2826S} relation, implying that the SED shapes produced are in line with those expected from the literature.  We note the presence of the X-ray weak quasar PG 1411+442, exhibiting a substantially lower $\alpha_{OX}$ than the rest of the population.

\subsection{Results for Individual AGN}
\label{individualagn}

\subsubsection{IC 4329A}

Inspection of images from NED reveal that the host galaxy for this AGN is an edge-on spiral galaxy.  As a result, the optical and UV continuum will be heavily contaminated with light from the galaxy and reddened by the pronounced dust lane visible in the optical images.  The OM points for this object do not produce a plausible fit to the \textsc{diskpn} model at all, and since the mass estimate from \cite{2004ApJ...613..682P} has very large errors, we exclude this object from our SED sample altogether.

\subsubsection{NGC 3227}

This AGN is classed as a Seyfert 1.5 in NED.  As already mentioned, it is also known to have significant optical and UV reddening \citep{2001ApJ...555..633C} and our SED fit to the OM data confirms this.  For the observation designated NGC 3227 (2), the point in the UVM2 band ($\mathrm{2310\AA}$) is lower in ${\nu}F_{\nu}$ than the UVW1 point ($\mathrm{2910\AA}$), contrary to the increasing shape expected from the UV disc emission, implying that the far UV has been significantly reddened.  The nucleus also seems to exhibit considerable variability in the UVW1 band, as the flux at this energy in observation (1) is a factor of $\sim2$ larger than that for observation (2). VF07 adopt a de-redenned optical--UV continuum shape following \cite{2001ApJ...555..633C}, but we leave the OM points as they are here. Additionally, this object is known to have significant and possibly variable X-ray absorption from \cite{2003A&A...397..883G}, who analyse one of the same sets of XMM observations used here.  Our simple \textsc{zwabs} model for the intrinsic absorption yields values of log($N_{H}/\rm{cm}^{-2}$) between $\sim$21.1--22.6, broadly consistent with the estimates of $\sim$21.3--22.0 from \cite{2003A&A...397..883G}.The presence of variable absorption presents difficulties in constraining the intrinsic X-ray luminosity.  Our value of $L_{\mathrm{2-10keV}} \approx 3.2 \times 10^{41}$ erg $\mathrm{s^{-1}}$ is comparable to the value $\sim2.2 \times 10^{41}$ erg $\mathrm{s^{-1}}$ from the more detailed modelling of \cite{2003A&A...397..883G}.  The two archival XMM observations used here were separated by six years, and the SED shape and derived parameters do not change appreciably between the two, as seen in Figs. \ref{bcvsedd} and \ref{seds}.

\subsubsection{NGC 3783}

This object is known to have an outflowing warm absorber \citep{2001ApJ...554..216K}, as discussed in detail by \cite{2004ApJ...602..648R}, and we obtain a value of log($N_{\rm{H}}/{\rm{cm}^{-2}}$)$\sim$ 21.3--21.4 from our simple \textsc{zwabs} model fit.  This differs significantly from the value $21.7$ in \cite{2004ApJ...602..648R}, presumably because of their more detailed model fit to the data. These authors propose that the warm absorber is relatively stable between multiple observations, implying that variations in the SED shape are more likely to be due to intrinsic continuum variation.  The two observations used here (separated by two days) reveal very similar SED continuum shapes and the spread in bolometric corrections obtained is small (a factor $\sim1.2$).

\subsubsection{NGC 4151}

The host galaxy for this object is a barred spiral galaxy.  It is also known to host high absorption and possesses a significant outflow \citep{2005ApJ...633..693K}.  This presents similar difficulties as discussed for the above objects for determining the intrinsic X-ray luminosity.  We have processed three observations spanning three days for this object, which reveal a comparatively large spread in bolometric correction.  Our simple \textsc{zwabs} model is probably insufficient to account for the complex nature of the outflowing absorption, and the underlying SED continuum could be more steady than suggested by the variable bolometric corrections.  More detailed modelling of this AGN may yield a more constant SED shape.

\subsubsection{PG 1411+442}

This PG quasar is known to be X-ray weak as discussed by \cite{2004A&A...414..107B}, with high intrinsic absorption (these authors find log($N_{H}/\rm{cm^{-2}}$)$\sim23.3$).  The bolometric correction determined for this object is $\sim1700$, but is likely to be significantly overestimated if the intrinsic luminosity has not been recovered accurately.  In VF07, another X-ray weak object was reported, PG 1011-040, but this object did not show signs of heavy intrinsic absorption according to previous studies \citep{2001ApJ...546..795G}.  Whether their low X-ray luminosities are intrinsic or not, the X-ray weak class of AGN, although small in number, continues to present a problem for the determination of a more universal AGN SED shape.

\subsubsection{NGC 4051}

We present results for two observations for this Narrow Line Seyfert 1 (NLS1) AGN, separated by a year and a half.  These two observations were analysed in detail by \cite{2006MNRAS.368..903P}, who note the large difference in X-ray flux between the two observations, distinctly harder spectrum for the low flux observation, and presence of a warm absorber in the spectrum. Other studies have sought to understand the source of the X-ray spectral variability seen in both high and low flux states \citep{2004MNRAS.347.1345U}, the detailed physics and geometry of the warm absorber seen in the long observation \citep{2008RMxAC..32..123K} and the complex emission and absorption line systems seen by \cite{2004ApJ...606..151O} in the long observation. Mechanisms which have been suggested for the X-ray spectral variability in this source are variable absorption along the line of sight (partial covering), spectral pivoting, or the presence of a reflection component from the accretion disc.  The analysis of \cite{2006MNRAS.368..903P} favours the last interpretation, and argues that the warm absorber, distinctly seen in the low-flux observation, is present during both observations.  If the large flux variations are due to underlying continuum variations, then the large spread we see in bolometric corrections (a factor $\sim5$) could be typical for such objects.  However, it seems that recovering the intrinsic luminosity is sufficiently complex here that the SED shape is again difficult to constrain.

\subsubsection{Mrk 335 and PG 1211+143}

These objects are both NLS1 AGN.  We present two observations for each of these objects, separated by almost seven years in the case of Mrk 335 and three years for PG 1211+143.   The analysis of the observations used in this study and the previous analysis in VF07 both place them well into the `high' Eddington regime ($\lambda_{\rm{Edd}}>\sim0.1$), with X-ray spectra which resemble the `soft' states of GBHs.  The 20 ks observation of Mrk 335 is part of the Target of Opportunity programme in which XMM observations were obtained contemporaneously with Swift observations.  This captures Mrk 335 at a remarkably low flux state \citep{2008arXiv0803.2516G}.

For high Eddington objects, it is likely that complex outflows may play a role in blurring our view of the nuclear emission.  It is possible to model the spectrum with partial covering or strong reflection due to light bending \citep{2004MNRAS.349.1435M} for both of these objects, again making the intrinsic X-ray luminosity difficult to determine.  This would naturally explain the large spread seen in the bolometric corrections (a factor $\sim2.9$ for Mrk 335; a factor $\sim3.7$ for PG 1211+143).  If the intrinsic nuclear X-ray luminosity is actually higher, this would tend to move the objects to higher Eddington ratios and smaller bolometric corrections.

\subsection{Revisiting bolometric corrections}
\label{revisit_bolcor}

\begin{figure*}
    \includegraphics[width=15cm]{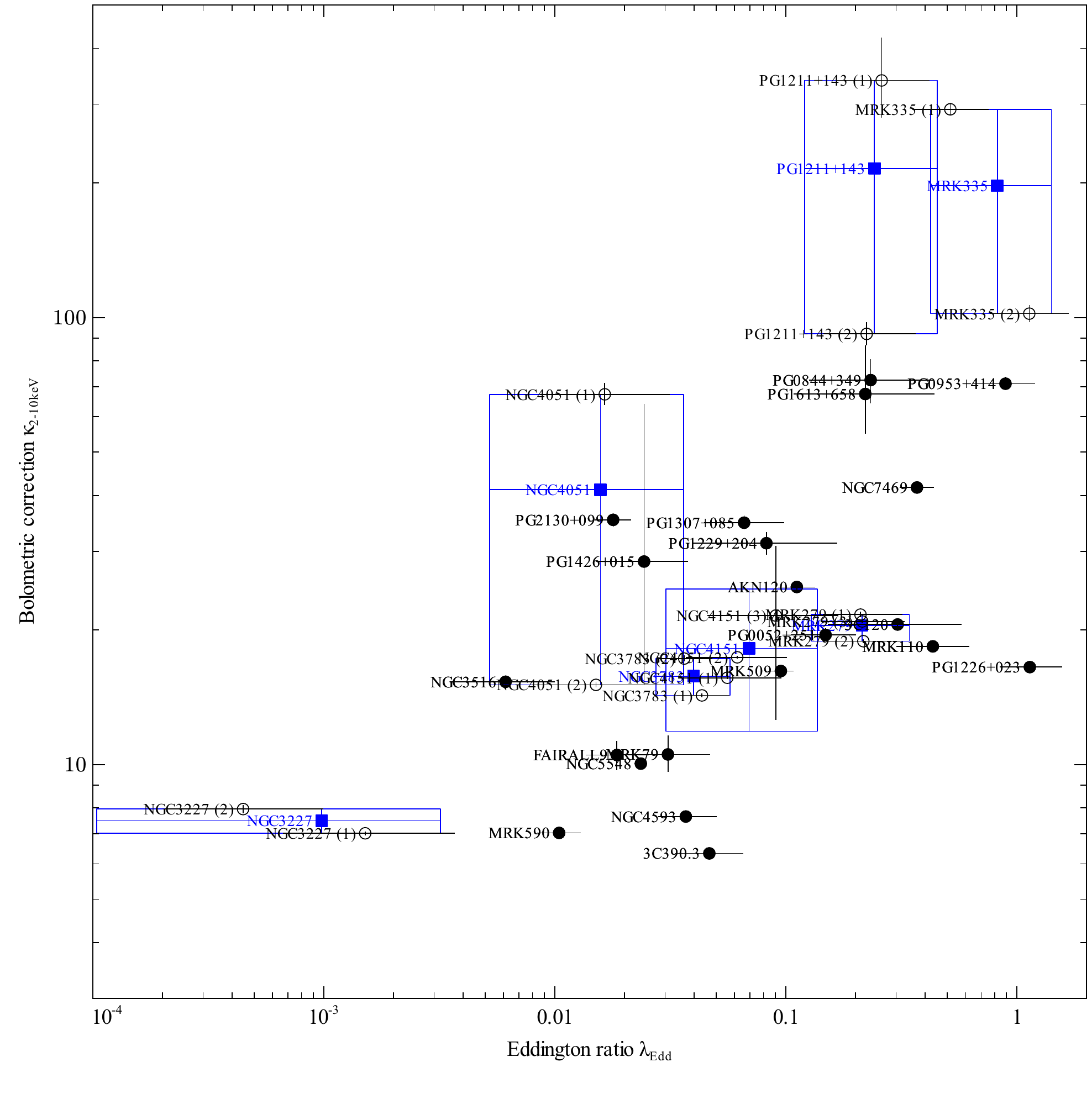}
    \caption{Hard X-ray (2--10keV) bolometric correction against Eddington ratio for the sample. Black filled circles represent objects where only one observation was used.  Empty circles represent the results from individual observations for objects with multiple observations.  Blue filled squares represent the average values determined from multiple observations, and the blue `bar \& box' error bars show the ranges spanned by multiple measurements, or the intrinsic error in the measurement if this is greater.}
\label{bcvsedd}
\end{figure*}

The simultaneous SEDs produced from our refined approach allow us to revisit the dependence of bolometric correction on Eddington ratio put forward in VF07, and the 2--10 keV bolometric corrections from the new SEDs are plotted against Eddington ratio in Fig. \ref{bcvsedd} (excluding the anomalous high bolometric correction for the X-ray weak AGN PG 1411+442).  We note that nuclear emission from radio loud objects is likely to be contaminated with jet emission.  We employ the values of radio loudness ($R = L_{\mathrm{6cm}}/L_{\mathrm{B}}$)  presented in \cite{2000ApJ...544L..91N} for this sample of AGN to identify three AGN (3C 120, 3C 390.3, PG 1226+023 $\equiv$ 3C 273) with $R>10$.  

\begin{figure*}
    \includegraphics[width=8.5cm]{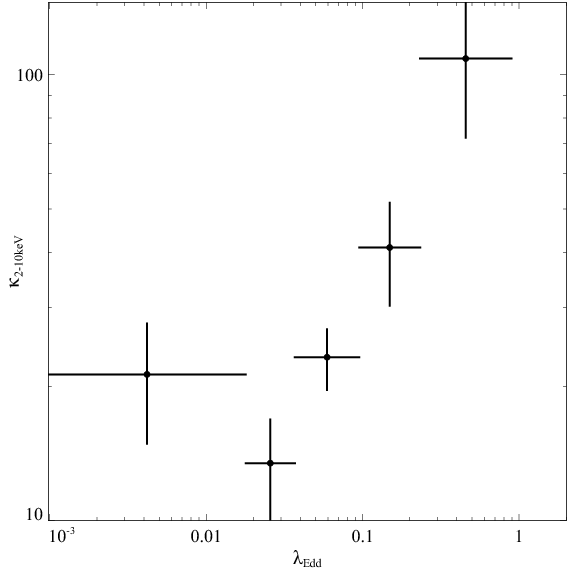}
    \includegraphics[width=8.5cm]{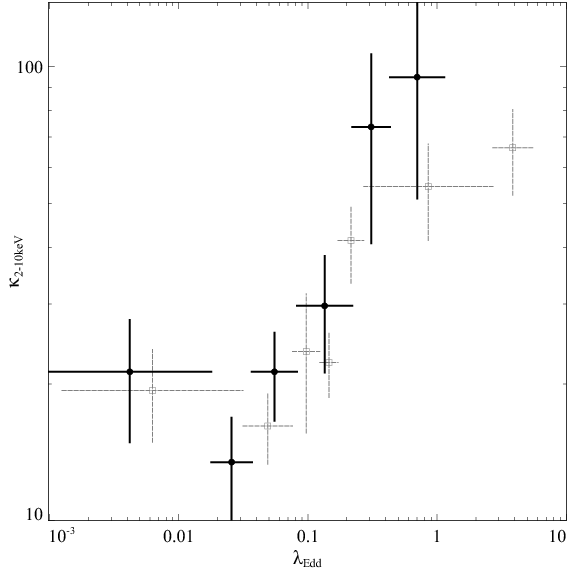}
    \caption{Left panel: Binned hard X-ray (2--10keV) bolometric correction against Eddington ratio for the sample; 5 objects per bin, 3 radio loud objects removed (see text for details).  Right panel: binned bolometric corrections against Eddington ratio including radio loud objects (5 objects per bin, the highest Eddington ratio bin has only 3 objects), presented with VF07 results (open grey squares and dotted grey lines) for comparison.  Error bars in bolometric correction are standard errors on the mean values of $\kappa_{\rm{2-10keV}}$, calculated from averaging bolometric corrections in each bin.}
\label{bcvsedd_binned}
\end{figure*}

We present the binned bolometric corrections against Eddington ratio with the radio loud objects removed in the left panel of Fig. \ref{bcvsedd_binned}.  Error bars on the mean bolometric correction values for each bin are standard errors on the means obtained when averaging the bolometric corrections.  For objects where multiple observations were available, the average bolometric correction and Eddington ratio calculated from all available observations were employed to calculate the bin average.  It should be noted that in some cases, the dramatically different bolometric corrections seen for the same object in different observations could be reflective of an underlying change of state of the object, and in such a scenario a simple average bolometric correction may not be meaningful.  The effect of partial covering or light bending on the bolometric correction of an object are such that objects which intrinsically have values ($\lambda_{i},\kappa_{i}$) will appear at ($\lambda_{0},\kappa_{0}$) where $\lambda_{0}<\lambda_{i}$ and $\kappa_{0}>\kappa_{i}$.

The right panel of Fig \ref{bcvsedd_binned} shows the same results with the radio loud objects included, for comparison with Fig. 12 of VF07.  With the consistent mass estimates and simultaneous SEDs presented here, we firstly confirm the previously identified trend for increasing bolometric correction with Eddington ratio.  However there is a marked shift towards larger bolometric corrections at all Eddington ratios.  This could be because the band-pass of the XMM-OM is less significantly affected by intrinsic AGN reddening than the \emph{FUSE} band-pass employed in VF07, giving larger bolometric luminosities.   

Our lowest Eddington bin could be subject to significant uncertainties, particularly because it includes NGC 3227 which is known to have significant reddening \citep{2001ApJ...555..633C}.  However, the increase in bolometric correction due to intrinsic reddening is likely to be as much as a factor of $\sim2$ (see VF07) which would only increase the average for this bin by a few per cent from the uncorrected value of $\sim22$.

\begin{figure}
    \includegraphics[width=8.5cm]{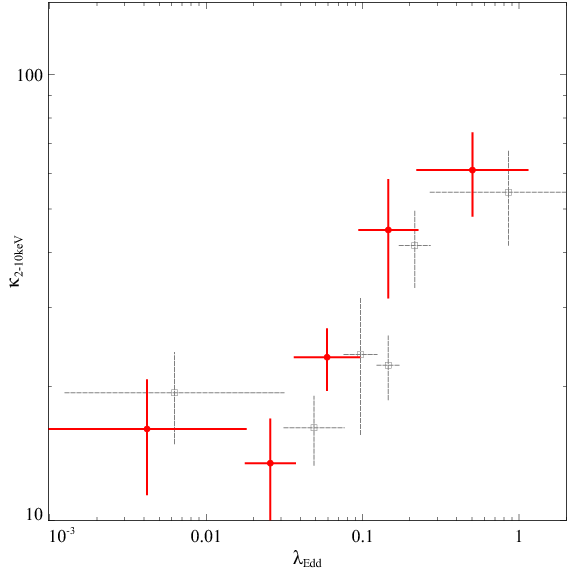}
    \caption{Binned hard X-ray (2--10keV) bolometric correction against Eddington ratio for the sample with radio loud objects and low X-ray flux observations (for NGC 4051, PG 1211+143 and Mrk 335) removed (red filled circles and lines). The VF07 results are presented for comparison (open grey squares and dotted grey lines).  All other conventions are as specified for Fig. \ref{bcvsedd_binned}}
\label{bcvsedd_binned_noRL_nolowxflux}
\end{figure}

We note from Fig. \ref{bcvsedd} that for many of the objects with multiple observations where variable absorption or partial covering have been reported, there is a possibility that the low X-ray flux observations are likely to give luminosities in the X-ray band significantly smaller than the intrinsic luminosity.  The degree of variation in SED shape is particularly large for NGC 4051, PG 1211+143 and Mrk 335 and so we plot bolometric corrections against Eddington ratio with the lower X-ray flux observations NGC 4051 (1), PG 1211+143 (1) and Mrk 335 (1) excluded from the calculations of average bolometric corrections (Fig. \ref{bcvsedd_binned_noRL_nolowxflux}).  The very high average bolometric corrections seen in the high Eddinton ratio bins in Fig. \ref{bcvsedd_binned} are now reduced substantially, reproducing the shape of the relation seen in VF07 more closely.

\subsection{Long term optical--UV variability}

The study of \cite{2007MNRAS.375.1479S} demonstrates that the rms variability of the optical--UV continua in AGN is of the order of a few per cent, over the course of 40 - 350 ks observations.  It is also instructive to compare the differences between optical--UV continua over longer time periods. There exist extensive studies on continuum and emission line variability in AGN by the `International AGN watch' consortium\footnote{http://www.astronomy.ohio-state.edu/~agnwatch/} using International Ultraviolet Explorer (\emph{IUE}), HST and ground based data.  For example, observations of NGC 3783 presented by \cite{1994ApJ...425..582R} show variations in flux of $\sim$30 per cent at 2700$\rm{\AA}$ over an 8 month period, and even larger factors are seen over longer periods: a factor of $\sim$2 in the 5340$\rm{\AA}$ flux for Fairall 9 (observed over $\sim$3 months -- \citealt{1997ApJS..112..271S}), and a factor of $\sim$7 in 5100 the $\rm{\AA}$ continuum flux for NGC 5548 (observed over eight years -- \citealt{1999ApJ...510..659P}).  Here, we present the variations in OM photometry for the three objects with biggest spread in 2--10keV bolometric corection, namely Mrk 335, NGC 4051 and PG 1211+143, to understand how this variability might affect the optical--UV SED shape and luminosities calculated from the SED.  We see from table \ref{table:omvar} that changes of the order of $\sim20$ per cent can introduce changes of up to a factor of $\sim2$ in the total UV disk flux, since small changes in flux in the optical--UV photometry will be extrapolated to larger changes over the whole energy range of the BBB.  While the variation in hard X-ray bolometric correction for an individual object is still attributable to X-ray variability in the main, we see from this that the variation in the optical--UV continuum can be a very significant contributor as well.  This supports the case for employing simultaneous X-ray and optical--UV data wherever possible in determining the total energy budget of AGN.

\begin{table*}
\begin{tabular}{p{1.8cm}|p{1.5cm}|p{1.5cm}|p{1.7cm}|p{1.7cm}|p{1.7cm}|p{1.7cm}|p{1.8cm}}
\hline
AGN&${\Delta}m(\rm{V})$&${\Delta}m(\rm{B})$&${\Delta}m(\rm{U})$&${\Delta}m(\rm{UVW1})$&${\Delta}m(\rm{UVM2})$&${\Delta}m(\rm{UVW2})$&$F_{\rm{disk}}^{\rm{(max)}}/F_{\rm{disk}}^{\rm{(min)}}$\\
\hline
Mrk 335&$--$&$0.090$&$0.166$&$--$&$0.248$&$--$&$1.92$\\
NGC 4051&$--$&$--$&$--$&$0.227$&$0.181$&$0.200$&$1.60$\\
PG 1211+143&$--$&$--$&$0.074$&$0.0642$&$0.038$&$0.003$&$1.03$\\
\hline
\end{tabular}
\caption{\label{table:omvar}Estimating the effect of long-term optical--UV variability on the total disk flux from \textsc{diskpn} (defined here as $F_{\mathrm{0.001-0.1keV}}$).}
\end{table*}

\subsection{Comparison with VF07}

\begin{figure*}
    \includegraphics[width=6cm]{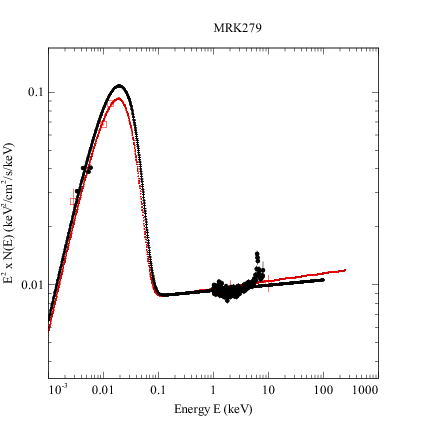}
    \includegraphics[width=6cm]{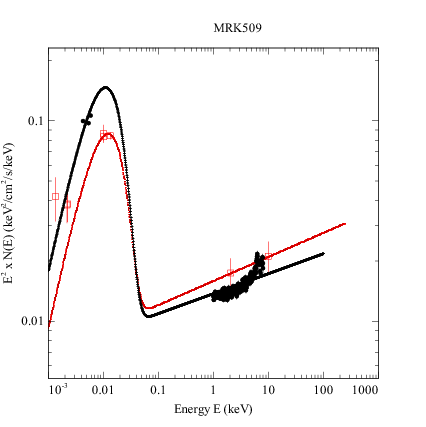}
    \includegraphics[width=6cm]{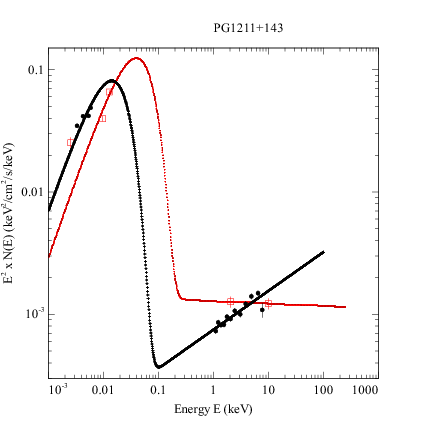}
    \includegraphics[width=6cm]{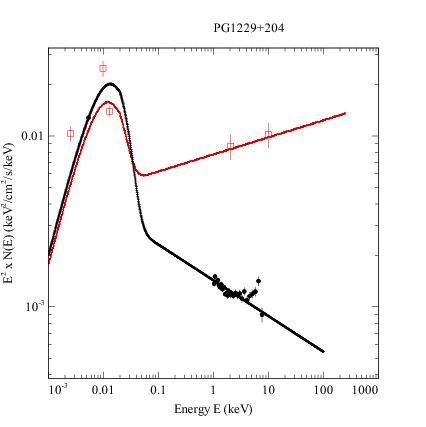}
  \caption{Comparisons of SEDs from this work (data: filled black circles, model: black crosses) with a selection from VF07 (data: open red squares, model: red dots).}
\label{sedscomparevf07}
\end{figure*}

\begin{figure}
    \includegraphics[width=9cm]{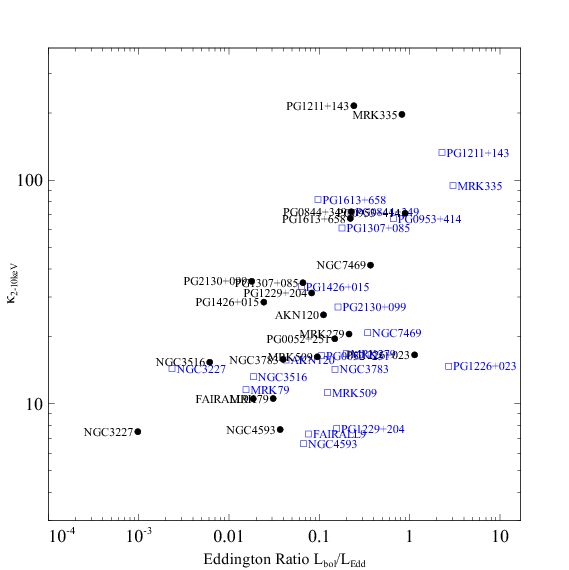}
    \caption{Comparison of bolometric corrections against Eddington ratios from the current work with the results from VF07.  Black filled circles represent results from this work; blue empty squares represent results from VF07.  Error bars are omitted for clarity.  For the typical size of error bars, refer to Fig. \ref{bcvsedd}}
\label{bcvseddcomparevf07}
\end{figure}

We note that 21 out of the 29 objects in this study were included in the previous study of VF07.  This allows a comparison of the simultaneous SEDs with the non-simultaneous SEDs from VF07, where the UV data were taken in all but a few cases from \emph{FUSE} observations of \cite{2004ApJ...615..135S}.  Firstly, this allows us to re-assess the importance of simultaneous observations, and secondly, to understand how significant reddening may affect the SED shapes in the far-UV \emph{FUSE} and \emph{XMM-OM} bandpasses.  We present a comparison of a sample of SEDs from VF07 with the SEDs from this work in Fig. \ref{sedscomparevf07}, and the hard X-ray bolometric corrections are compared in Fig. \ref{bcvseddcomparevf07}.  We remove an erroneous optical SED point for the AGN NGC 3783, which was incorrectly named as NGC 3873 in VF07 (this alteration does not alter the previously published bolometric correction).

We note that the range of Eddington ratio and bolometric corrections obtained here is very similar to those obtained in VF07.  At high Eddington ratio, the values for Eddington ratio from VF07 tend to exceed those obtained in this study.  This could be in part due to the masses being systematically smaller in VF07.  The masses for the overlapping sources in VF07 were obtained from the previous reverberation mapping catalogue of \cite{2000ApJ...533..631K} and the single-epoch virial mass measurements of \cite{2005MNRAS.356.1029B}, along with a few objects with masses from \cite{2004ApJ...613..682P} (NGC 3227, NGC 3516, NGC 4593, Mrk 279). Eddington ratios for individual objects in this study range from a factors of $\sim0.1$ to $\sim2.6$ of their values in VF07, and on average are a factor $\sim0.8$ of the VF07 values.  The \cite{2004ApJ...613..682P} masses are generally larger than the values used in VF07, but the differences in Eddington ratio between the two studies can be ascribed equally to differences in black hole mass estimates and bolometric luminosities.  Bolometric corrections from this study range from factors of $\sim0.5$ to $\sim4.0$ of the VF07 values, and on average are slightly larger in this study (by a factor $\sim1.34$).  We note that if NGC 3227 is excluded from this average (because the obvious presence of reddening in the spectrum used here prevents a meaningful comparison with the reddening-corrected spectrum used in VF07), we find that the bolometric corrections in this work are on average around $\sim1.38$ times larger than the VF07 values.

\subsection{Average SEDs for high and low Eddington ratios, comparisons with a real accretion disc SED}

\begin{figure}
    \includegraphics[width=8.5cm]{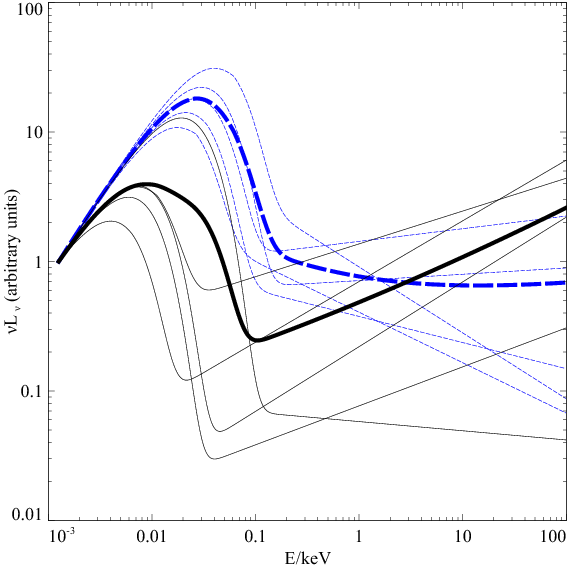}
    \caption{Average SEDs for AGN in the highest ($\langle\lambda_{\rm{Edd}}\rangle = 0.61$, thick blue dashed line) and lowest ($\langle\lambda_{\rm{Edd}}\rangle = 0.01$, thick black solid line) Eddington ratio bins.  The individual SEDs for AGN in those bins are depicted using thin blue dashed lines ($\langle\lambda_{\rm{Edd}}\rangle = 0.61$ bin), and thin black solid lines ($\langle\lambda_{\rm{Edd}}\rangle = 0.01$ bin).}
\label{averageseds}
\end{figure}

We present the rest-frame average SEDs (normalised at 1eV) for the AGN in the highest and lowest Eddington ratio bins in order to compare the SED shape characteristics (Fig. \ref{averageseds}).  We adopt the higher X-ray flux observations where multiple observations were available, for the reasons outlined in section \ref{revisit_bolcor}.  The highest Eddington ratio bin has $\langle\lambda_{\rm{Edd}}\rangle = 0.61$ and we note the presence of a substantially larger accretion disc component, coupled with a softer X-ray spectrum than the average SED for the low Eddington ratio bin ($\langle\lambda_{\rm{Edd}}\rangle = 0.01$).  This confirms the trends reported in VF07 for the variation in SED shape with Eddington ratio.  The average high and low Eddington ratio SEDs have markedly different fractions of ionizing UV luminosity ($L_{\rm{13.6-100eV}}/L_{\rm{bol}}$).  The average low Eddington ratio SED in this study has $L_{\rm{13.6-100eV}}/L_{\rm{bol}}\sim21$ per cent, rising to $\sim59$ per cent for high Eddington ratio (taking $L_{\rm{bol}}=L_{\rm{0.001-100keV}}$).   The average high and low Eddington ratio SEDs in VF07 display more pronounced differences; at low Eddington ratios the fraction is typically $\sim9$ per cent, rising to $\sim54$ per cent for high Eddington ratios; but the less dramatic change seen in this work is expected because the range of Eddington ratios spanned here is smaller than that in VF07 (the central Eddington ratios of the high and low Eddington bins in VF07 span a factor of $\sim140$; here they span a factor of only $\sim50$).

\begin{figure}
    \includegraphics[width=8.5cm]{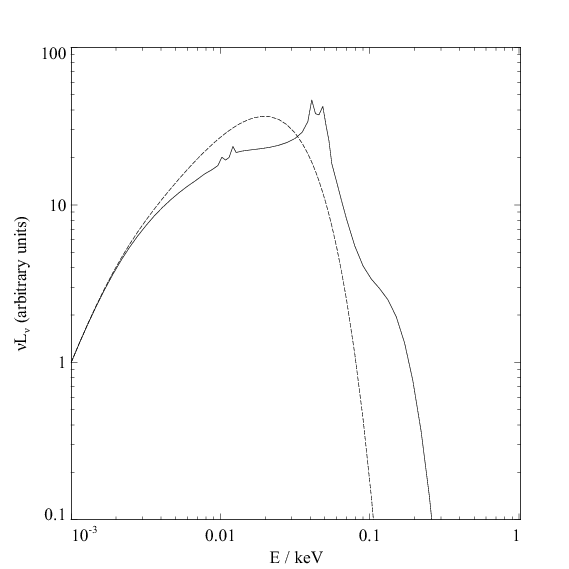}
    \caption{Comparison between a more realistic accretion disc SED with the equivalent black-body spectrum.  The curves are taken from Fig. 4(b) from \protect\cite{1992MNRAS.258..189R} with the authors' permission.}
\label{realdisksed}
\end{figure}

In this work, we adopt a simple multicolour disc model (\textsc{diskpn}) for the BBB.  It is instructive to compare how the optical--UV data compare with more sophisticated accretion disc models.  \cite{1992MNRAS.258..189R} present self-consistent calculations for Hydrogen-Helium accretion disc spectra in AGN, based on the standard $\alpha$-disc model \protect\citep{1973A&A....24..337S}, including features such as the H-\textsc{I} Lyman edge and He-\textsc{II} Lyman $\alpha$ and $\beta$ emission lines.  We reproduce the disc SED for an accretion disc surrounding a $M_{\rm{BH}}=10^{8}M_{\bigodot}$ black hole with Eddington ratio $\lambda_{\rm{Edd}}=0.167$, along with the equivalent black-body disc spectrum for comparison. (Fig. \ref{realdisksed} reproduced from Figure 4b of their paper, with the authors' consent).  The `real' disc spectrum follows the equivalent black-body spectrum at low energies ($E\lesssim 4$ eV) but deviates at higher energies, where the H-\textsc{I} and He-\textsc{II} features are evident.  The descending high-energy tail is shifted significantly from that for the equivalent black-body disc spectrum.  However, since the H-\textsc{I} and He-\textsc{II} features are in the unobservable extreme-UV, simple black-body model used here is adequate for our purposes, since only near-UV data are available to us.  The difference in total flux between the `real' disc model and the blackbody is negligible (of the order of $\sim5$ per cent) and would not significantly alter bolometric luminosities.  For a handful of sources in this work, the UV points are close to the turnover in a standard black-body disc spectrum.  In these cases, the data points would lie at most a factor of $\sim2$ above the equivalent `real' disc spectrum, since the model values differ by approximately this factor at the turnover.  If a `real' disc spectrum were fit to these points successfully, the resulting disc luminosities would increase by a comparable amount.  This could only be significant for perhaps 6 objects (3C 390.3, Fairall 9, NGC 3516, PG 1307+085, PG 1426+015, PG 2130+099) out of the 29 in the sample.  We exclude NGC 3227 from this list because the low UV flux is already attributed to reddening.

\section{Systematics}

\subsection{Biases in the sample}

We note the possibility that the increase in bolometric correction with Eddington ratio could be influenced by the presence of $L_{\rm{bol}}$ on both axes.  We assess the importance of this by generating a random sample of AGN with X-ray luminosities, redshifts and masses randomly chosen within the ranges shown in this sample.  A conservative flux limit of $10^{-15}\rm{ergs^{-1}cm^{-2}}$ was employed in choosing luminosities.  A random bolometric correction is assigned to each AGN between 5 and 1000, ensuring that the resulting bolometric luminosity does not lie outside the limits in our sample.  A plot of bolometric correction against Eddington ratio calculated from these assumptions will be constrained to lie within certain limits, in particular restricting the allowed bolometric corrections to lower values at very low Eddington ratios.  However, the range of scatter seen in the points is far greater than that seen in the data, implying that the limits imposed by plotting $L_{\rm{bol}}/L_{\rm{2-10keV}}$ against $L_{\rm{bol}}/L_{\rm{Edd}}$ are not relevant here.  The gradient implied from the random sample is also steeper than that seen in the data.  If a correlation is introduced between $M_{\rm{BH}}$ and $L_{\rm{2-10keV}}$, the limits imposed by the data are narrowed somewhat, but such a correlation has not been reported, either in local or deeper samples (\citealt{2005ApJ...624..155P}, \citealt{2007A&A...474..755B}).  These considerations suggest that the changes in SED shape/bolometric correction with Eddington ratio are intrinsic.

\subsection{Corrections to reverberation mapping masses due to radiation pressure}

Reverberation mapping masses have previously known to have an uncertainty of a factor $\sim3$.  However if this is symmetric, it is unlikely to change the trend of the bolometric correction--Eddington ratio plot significantly.  However, recent work by \cite{2008arXiv0802.2021M} discusses the need to take into account the effect of radiation pressure when calculating virial mass estimates.  They argue that in calculating the size of the broad line region (BLR) required for virial BH mass estimates, the effect of radiation pressure on the BLR clouds must be included as well as gravity.  This becomes especially important in AGN where the assumption that the system is gravitationally bound is likely to be violated, such as Narrow Line Seyfert 1 AGN (NLS1s) which are thought to be accreting at high, possibly super-Eddington ratios.  Their refined expression for the BH mass includes the standard virial term with the geometry-dependent scale factor $f$, and an additional term dependent on the radiation pressure scaled by a new factor $g$, which is dependent on the density of the BLR clouds and the SED shape of the illuminating central source.  They provide the empirically calibrated correction for radiation pressure in equation (6) of their paper, as a function of the 5100$\rm{\AA}$ monochromatic luminosity ($\lambda L_{\rm{\lambda}}(\rm{5100\AA})$).

We try to assess the importance of this effect on Eddington ratios and bolometric corrections by applying a first order correction to the reverberation mass values using the 5100 $\rm{\AA}$ luminosities from our AGN SEDs.  We adopt the values $f=3.1$ and $\rm{log}(g)=7.6$ as found by \cite{2008arXiv0802.2021M} from requiring that the reverberation masses lie on the $M_{BH}-\sigma$ relation. The average value of $\mathrm{M_{BH}^{new}/M_{BH}^{old}}$ is $\sim1.35$ for the whole sample.  \cite{2008arXiv0802.2021M} calculate $\lambda L_{\rm{\lambda}}(\rm{5100\AA})$ more accurately by subtracting the host galaxy following the approach of \cite{2006ApJ...644..133B}, and find a similar average value for the whole sample, $\mathrm{M_{BH}^{new}/M_{BH}^{old}\sim1.25}$ (obtained via private communication with the author).  We do not find a systematically higher value for the four known NLS1s in the sample (NGC 4051, Mrk 335, PG 1211+143 and PG 2130+099), and some of the low Eddington ratio AGN have smaller masses with the new calculation, because of the lower value of $f$ used (\citealt{2004ApJ...613..682P} use $f=5.5$).

\begin{figure}
    \includegraphics[width=8.5cm]{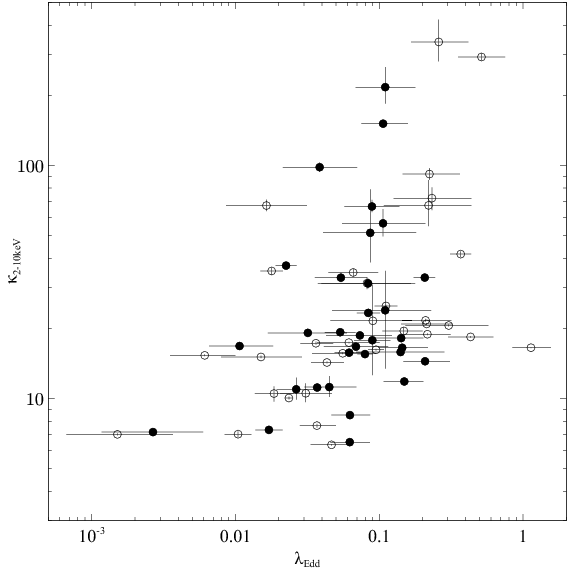}
    \caption{Hard X-ray (2--10keV) bolometric corrections against Eddington ratio with masses corrected for radiation pressure according to \protect\cite{2008arXiv0802.2021M} (filled circles), compared to results using the original reverberation mapping $M_{BH}$ estimates (empty circles).}
\label{bcvsedd_massup}
\end{figure}

We plot the bolometric corrections against Eddington ratio for the objects with individually corrected masses in Fig. \ref{bcvsedd_massup}, eliminating one object where the new mass value (and resultant \textsc{diskpn} normalisation) results in an unphysical fit. The Eddington rates and bolometric corrections before correction are plotted for comparison, and as expected, since the high Eddington ratio objects are those in which radiation pressure is likely to be higher, their corrected masses result in Eddington ratios lowered by up to $\sim1$ dex with bolometric corrections changing by $\sim0.1-0.4$ dex.  Some of the revised mass estimates for low Eddington ratio AGN are now smaller, causing the lower Eddington ratio objects to shift towards higher ratios, revealing what may be an intrinsically narrower distribution of Eddington ratios in the sample.  Overall, the range in bolometric corrections seen is not altered significantly.

\subsection{Probing the low Eddington ratio regime}

It is clear that this sample contains few sources with $\lambda_{Edd}\ll0.1$, apart from NGC 3227.  This was also identified as a problem with the \emph{FUSE} sample used in VF07, where supplementary UV data were gathered from other sources in the literature to augment the low luminosity/Eddington ratio regime.  The tendency for reddening to be significant in this regime also presents problems.  Additionally, it is known from \cite{2008MNRAS.385L..43F} that the majority of sources with significant absorption tend to cluster at low Eddington fractions.   These observations suggest that it is necessary to use the re-processed IR to understand the accretion emission in low luminosity Seyferts and AGN.  \protect\cite{2007A&A...468..603P} present precisely this type of analysis for eight high-redshift luminous obscured quasars, combining \emph{XMM-Newton} observations with \emph{Spitzer} IRAC, MIPS $24{\mu}m$ and $K_{S}$-band observations.  They remark on the likelihood that a significant proportion of luminous obscured AGN may escape identification due to faint optical counterparts, implying large obscuration and directing us naturally to the IR to study the reprocessed nuclear (torus) emission.  They estimate the bolometric luminosity by adding the total X-ray luminosity (0.5--500 keV) to the total IR luminosity (1-1000${\mu}$m) and thus calculate bolometric corrections for their sample.  Interestingly, they find $\mathrm{\kappa_{2-10keV}}\approx25$ on average for these AGN, which all seem to be in a low-Eddington fraction regime ($0.008\lesssim\lambda_{Edd}\lesssim0.170$).  This would support the hints from VF07 and this work that $\mathrm{\kappa_{2-10keV}}$ is low for low Eddington fraction AGN.  The distinctive accretion and SED characteristics of low-Eddington fraction AGN could be clarified by a similar analysis of a much larger sample of AGN using both X-ray and IR data.  We would then be provided with a window onto the accretion modes at work in both `unobscured' and `obscured' AGN.

\section{Summary and Conclusions}

We have presented, for the first time, a set of simultaneous optical--to--X-ray SEDs for 29 AGN from the \cite{2004ApJ...613..682P} reverberation-mapped sample.  By using simultaneous data and consistent mass estimates, this work represents a refinement over VF07, yet reinforces their main conclusions.  

\begin{itemize}

\item In particular, we confirm an increase in bolometric correction with Eddington ratio, with bolometric corrections of $\kappa_{\rm{2-10keV}}\sim15-30$ for $\lambda_{\rm{Edd}}\lesssim0.1$, $\kappa_{\rm{2-10keV}}\approx20-70$ for $0.1\lesssim\lambda_{\rm{Edd}}\lesssim0.2$ and $\kappa_{\rm{2-10keV}}\sim70-150$ for $\lambda_{\rm{Edd}}\gtrsim0.2$, excluding three significantly ($R>10$) radio loud objects.  We note that the bolometric corrections from this study are generally higher than those in VF07 by a factor of $\sim1.3-1.4$, and Eddington ratios tend to be marginally smaller in this study.  The change in bolometric correction is likely due to the reduced intrinsic reddening in the OM bandpass; this has the effect of increasing the overall optical--UV luminosity, giving higher bolometric corrections.  Here we demonstrate that the changes in the SED with accretion rate $\lambda_{\rm{Edd}}$ are consistent with their results when performed on a sample with fully re-reduced simultaneous data and consistent mass estimates.


\item The availability of multiple observations for individual objects pinpoints cases where constraining the intrinsic SED shape is problematic because of the complex and variable nature of intrinsic absorption, or the presence of outflows.  NGC 3227 shows significant UV reddening in one of the available observations, but little overall variation in the SED shape aside from this.  NGC 3783 and NGC 4151 both exhibit outflows, and for the latter in particular, this manifests as a relatively large srpead in bolometric correction (a factor of $\sim2$). PG 1411+442 is X-ray weak, but the large bolometric correction of $\sim1700$ is likely to be severely overestimated because of significant absorption.  Conversely, it is also known that the X-ray spectra of X-ray weak AGN are not all accountable for with large amounts of absorption, and so they remain a somewhat unknown quantity in the larger `puzzle' of the AGN SED.  The two observations of NGC 4051 in the XMM archive are known to have radically different X-ray fluxes and yield very different bolometric corrections, spanning a factor greater than $\sim5$.  A warm absorber, partial covering or a reflection component from the accretion disc could account for this large variation.  Mrk 335 and PG 1211+143 are both NLS1s with high Eddington fractions, where the presence of outflows may confuse our view of the accretion emission.  These considerations highlight that recovering the underlying continuum in AGN is a messy process and often requires a careful, object-specific approach.  

\item Relatively small changes in the optical--UV photometry over months to years can lead to comparatively more significant changes in the total extrapolated flux in the BBB energy range, reinforcing the preference for simultaneous X-ray and optical--UV observations where possible when calculating the total accretion energy budget.

\item Reverberation mapping provides a powerful and reliable method for getting SMBH mass estimates, which are a prerequisite for accurate Eddington ratio estimates.  However, the uncertainty in the precise geometry and kinematics of the BLR contributes a factor of $\sim3$ uncertainty to the resultant mass estimates, but recent work by \cite{2008arXiv0802.2021M} has highlighted a possible systematic shift towards larger RM masses if radiation pressure is taken into account.  We estimate that the ratio of the new to old masses for the \cite{2004ApJ...613..682P} sample is $\sim1.35$ on average, but with some low Eddington AGN actually exhibiting a decrease in mass.  This does not significanlty alter the spread of bolometric corrections but may narrow the distribution of Eddington ratios in the sample overall.

\item It is clear that the low Eddington ratio regime is difficult to constrain with relatively bright Seyfert 1 AGN and quasars.  The study of \cite{2008MNRAS.385L..43F} on deep samples such as the Chandra Deep Field--South and Lockman hole suggests that the majority of sources are of low Eddington ratios. Moreover, they are often significantly obscured, and therefore optical--UV monitoring of the accretion disc emission is likely to be plagued by significant reddening problems.  The work of \cite{2007A&A...468..603P} suggests that the bolometric corrections for low Eddington ratio AGN are likely to be small as obtained in VF07 and in this work, and a large sample of AGN with X-ray and IR data would then be useful for completing the picture of AGN accretion emission across a much larger range of Eddington ratios.

\item We have demonstrated that good quality simultaneous optical, UV and X-ray data, coupled with consistent mass measurements continue to provide opportunities for understanding the accretion characteristics of AGN in new ways, particularly for the brighter, unobscured objects.  The large archives from \emph{XMM-Newton} and \emph{Swift} provide great potential for further illuminating study.  If such work is coupled with a consistent, well calibrated mass estimation technique, we will also be able to calculate accretion rates for the AGN from accurate determinations of their bolometric luminosity.  Single-epoch virial mass measurements and various forms of the $M_{\rm{BH}}-L_{\rm{bulge}}$ relation offer two such avenues, and a large database of SEDs with accretion rates for a well-defined sample of AGN remains an interesting subject for future work.

\end{itemize}

\section{Acknowledgements}

RVV acknowledges support fom the Science and Technology Facilities Council (STFC) and ACF thanks the Royal Society for Support.  This research has made use of observations obtained with \emph{XMM-Newton}, an ESA science mission with instruments and contributions directly funded by ESA member states and the National Aeronautics and Space Administration (NASA).  We thank Craig Gordon and Keith Arnaud from the \textsc{xspec} developer team at NASA/Goddard Space Flight Center for quick responses to queries on the \textsc{xspec12} package and for promptly issuing updates to the software needed by this work.  We also thank the anonymous referee for helpful suggestions.  This research has also made use of the NASA Extragalactic Database (NED) which is operated by the Jet Propulsion Laboratory, California Institute of Technology, under contract with NASA.

\bibliographystyle{mnras} 
\bibliography{xmmagnseds_revmapmass}

\end{document}